\documentclass[3p,times,procedia,number]{elsarticle}
\usepackage{ecrc}
\usepackage{color}
\volume{00}
\firstpage{1}
\journalname{Procedia Engineering}
\runauth{J.-F. Remacle}
\jid{proeng}
 \biboptions{sort&compress}

\usepackage{hyperref}
\usepackage{amsmath,amstext,bm, mathtools}
\usepackage{makeidx}         
\usepackage{graphicx}        
\usepackage{multicol}        
\usepackage[bottom]{footmisc}
\usepackage{dsfont} 

\usepackage[ruled,vlined]{algorithm2e}
\usepackage{listings,color}

\usepackage{mathtools}
\DeclarePairedDelimiter{\ceil}{\lceil}{\rceil}

\newcommand{\pointSet}{\ensuremath{P}}
\newcommand{\DT}{\ensuremath{\mbox{DT}}}
\newcommand{\dd}{\ensuremath{\mbox{d}}}
\newcommand{\ttt}{\ensuremath{\mbox{t}}}
\newcommand{\mvx}{\ensuremath{\mathbf x}}
\newcommand{\mvp}{\ensuremath{\mathbf p}}
\newcommand{\mvc}{\ensuremath{\mathbf c}}
\newcommand{\reals}{\ensuremath{\mathds{R}}}

\graphicspath{{./}{../figures/}}

\makeindex             

\begin{document}
\begin{frontmatter}

\title{Fast and robust mesh generation on the sphere \\ Application to
coastal domains.}
\author[a]{ Jean-Fran{\c c}ois Remacle}
\author[a]{ Jonathan Lambrechts}
\address[a]{Universit\'e catholique de Louvain, Institute of Mechanics,
Materials and Civil Engineering (iMMC), B\^atiment Euler,
Avenue Georges Lema\^itre 4, 1348 Louvain-la-Neuve, Belgium}

\begin{abstract}
This paper presents a fast an robust mesh generation procedure that is
able to generate meshes of the earth system (ocean and continent) in
matters of seconds. Our algorithm takes as input a standard shape-file
i.e. geospatial vector data format for geographic information
system (GIS) software. The input is initially coarsened in order to
automatically remove unwanted channels that are under a desired
resolution. A valid non-overlapping 1D mesh is then created on the
sphere using the Euclidian coordinates system $x,y,z$. A modified Delaunay
kernel is then proposed that enables to generate meshes on
the sphere in a straightforward manner without parametrization. 
One of the main difficulty in dealing with geographical data is the
over-sampled nature of coastline representations. We propose here an
algorithm that automatically unrefines coastline data. Small features
are automatically removed while always keeping a valid
(non-overlapping) geometrical representation of the domain.
A Delaunay refinement procedure is subsequently applied to the
domain. The refinement scheme is also multi-threaded at a fine grain level,
allowing to generate about a million points per second on 8 threads.
Examples of meshes of the Baltic sea as well as of the global ocean
are presented.
\end{abstract}


\cortext[cor1]{Jean-Fran{\c c}ois Remacle.}
\end{frontmatter}

\section{Introduction}
Traditional ocean models are based on finite differences schemes on Cartesian grids \citep{griffies2000}. 
It is only recently that unstructured meshes have been used in 
ocean modeling \cite{icom,white2007a,danilov2005}, essentially using
finite elements. One of the advantages of unstructured grids 
is their ability to conform to coastlines.

As unstructured grid ocean models began to appear,
mesh generation algorithms were either specifically developed or
simply adapted from classical engineering tools. 
\cite{leprovost1994} use the mesh generation tools of 
\cite{henry1993} on several subdomains to obtain a mesh of the world ocean, aiming at global scale tidal modeling. 
Further, \cite{lyard2006} use a higher resolution version of the same kind of meshes with the state of the art FES2004 tidal model.
\cite{hagen2001} give two algorithms to generate meshes of coastal domains, and use them to model tides in the Gulf of Mexico.
\cite{legrand2006} show high-resolution meshes of the Great Barrier Reef (Australia).
At the global scale, \cite{legrand2000} and \cite{meshimperial} developed specific algorithms to obtain meshes of the world ocean.
More recently, we have developed a proper CAD model of ocean
geometries \cite{lambrechts2008multiscale}. This model relies on the
stereographic projection of the sphere which is conformal i.e. it
conserves angles. 
This approach has been quite successful
up to now: we and other teams have applied it to numerous coastal
domains \cite{sassi2011tidal,lambrechts2008multi}.

Our CAD approach has two major drawbacks. First, at least two maps
are required to cover the whole sphere, making it awkward for
atmosphere simulations for example. Then, using splines is maybe not
the most robust/natural manner  for describing
coastlines: geographical information systems provide description of
coastlines as series of non-overlapping closed polygons and using
splines may lead to intersections.

Here, a new approach that addresses both issues is proposed. 

A modified Delaunay kernel is first presented that allows to generate meshes on
the unit sphere.  Based on our recent paper \cite{remacle2015two}, a
multithreaded version of this new kernel has been
implemented  that allows to triangulate over one million points per
second on the sphere on a standard quad-core laptop. This new approach
does not rely on any parametrization and has all the proof
structure of the usual Delaunay kernel (proof of termination,
angle-optimality, polynomial complexity).

In this new approach, the most refined representation of coastlines
available in the geographical system is used as input . 
A constrained Delaunay mesh of the whole data set is created using 
the new Delaunay procedure. This first mesh
allows to automatically and robustly remove from the domain any
water channel that has a width that is smaller than a given threshold (this
threshold being possibly variable in space). This first step leads to a coarsened
version of the shapes where locally small features have been
removed. We show that our procedure produces a valid (non-overlapping)
boundary description of the domain. Then,  a one-dimensional mesh is created
using the coarse geometry.  Finally, a multi-threaded version of the
edge-based Delaunay refinement procedure of \cite{george1997improvements} has been
used to saturate the domain with points and triangles. 

The developments that are presented here have been released as a self
consisten open source code that can be used as a stand alone program
or that can be plugged in other softwares such as Gmsh \cite{geuzaine2009gmsh} or
QGIS \cite{qgis2011quantum}. 

\section{Delaunay triangulation on the sphere}
Here we consider the unit 3D sphere $S$ centered at the origin ${\bf o}(0,0,0)$: $S=\{\mvx(x,y,z) \in \reals^3~|~x^2+y^2+z^2 = 1\}$.
Any section of a sphere by a plane is a circle. We distinguish \emph{great circles} that are sections of a sphere that diameter is 
equal to the diameter of the sphere and  \emph{small circles}  that are any other section.

Consider two points $\mvp_1$ and $\mvp_2$ on the sphere. The shortest path between $\mvp_1$ and $\mvp_2$ is called the \emph{geodesic path}. 
It can be shown that geodesic paths on the sphere are segments of a great circle. The \emph{geodesic distance} between $\mvp_1$ and $\mvp_2$
is the length of the great circular arc joining $\mvp_1$ and $\mvp_2$. We call it $\dd(\mvp_1,\mvp_2)$.

\begin{figure}[h!]
 \begin{center}
 \begin{tabular}{cc}
 \scalebox{0.8}{\includegraphics{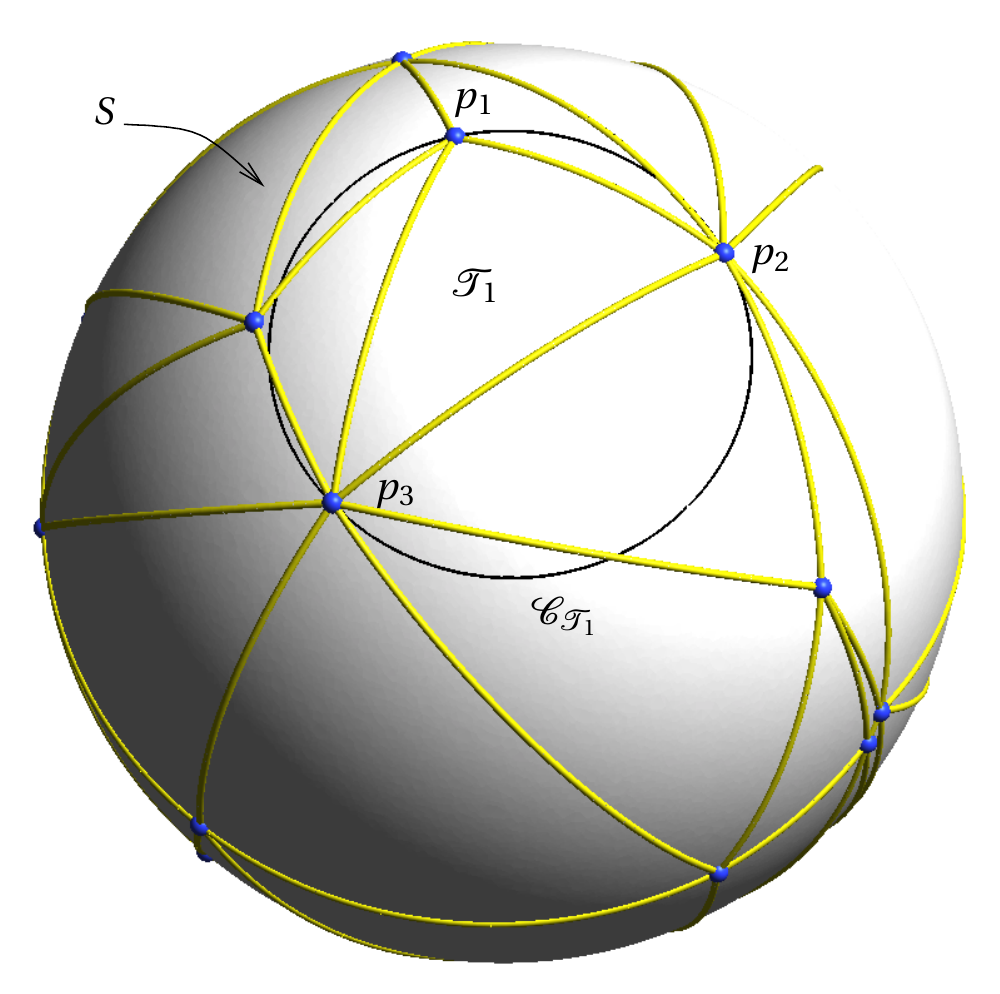}}&
 \scalebox{0.8}{\includegraphics{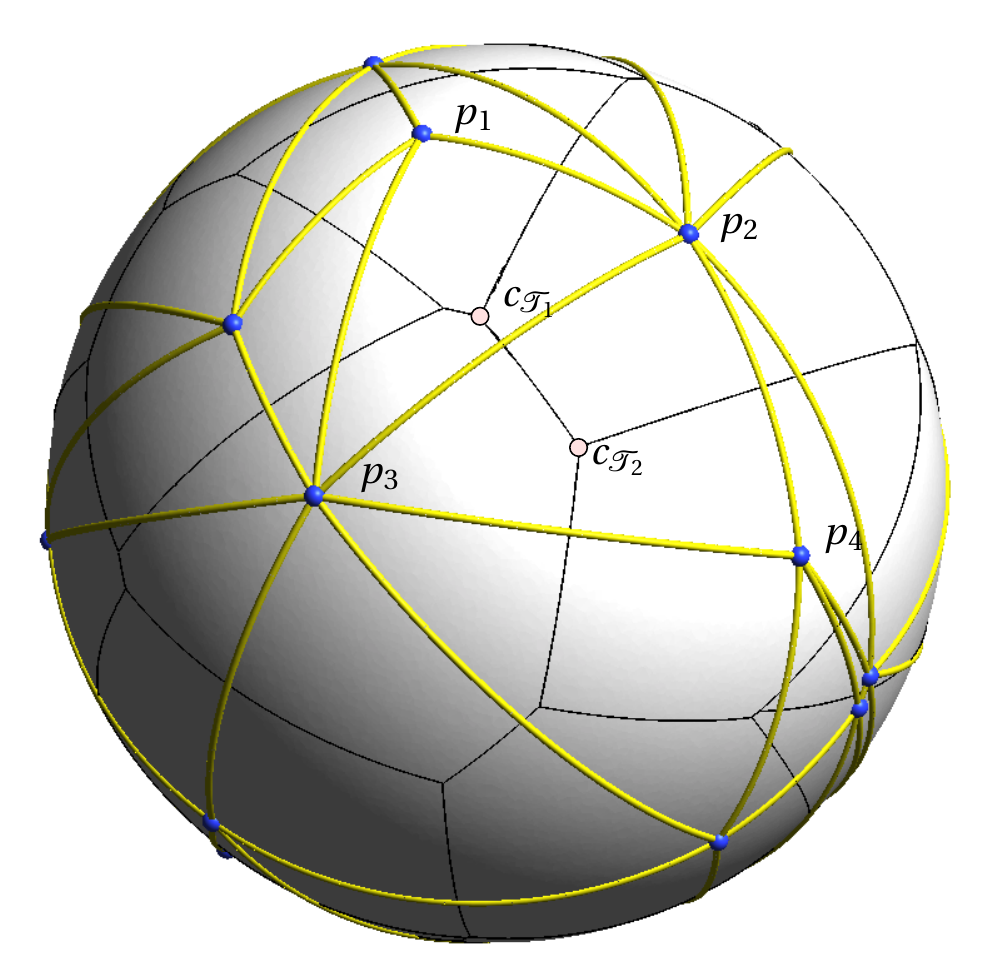}}
\end{tabular}
\end{center}
\caption{Delaunay triangulation on a sphere. Spherical triangle $\mathcal{T}
  (\mvp_1,\mvp_2,\mvp_3)$ is drawn with its spherical circumcircle
  $\mathcal{C}({\mathcal T})$  that is empty. Every triangle edge 
  (in yellow, like $\mvp_1\mvp_2$) is a great circle.\label{fig:spht}}
\end{figure}

A spherical triangle $\mathcal{T}_1 (\mvp_1,\mvp_2,\mvp_3)$ (see Figure \ref{fig:spht}) is a figure formed on the surface of a sphere by three great circular arcs intersecting pairwise in three vertices
$\mvp_1$, $\mvp_2$ and $\mvp_3$.  A spherical triangle is sometimes called an
Euler triangle. Spherical triangles have an orientation that is
computed as the sign of the volume $\|\mvp_1,\mvp_2,\mvp_3,{\bf o}\|$ of tetrahedon
$\ttt(\mvp_1,\mvp_2,\mvp_3,{\bf o})$ is positive, with ${\bf o}$ the center of $S$.

The circumcircle $\mathcal{C}_{\mathcal T_1}$ of the
spherical triangle ${\mathcal T}_1$ is the small circle  that
is formed by the section of $S$ by the plane defined by points $\mvp_1$,
$\mvp_2$ and $\mvp_3$ (see Figure \ref{fig:spht}). The circumcircle 
$\mathcal{C}_{\mathcal T_1}$ divides the sphere in two parts. Consider
a point $\mvp$ of $S$:
\begin{itemize}
\item $p$ is inside $\mathcal{C}_{\mathcal T_1}$ if
  $\|\mvp_1,\mvp_2,\mvp_3,\mvp\| < 0$.
\item $\mvp$ is outside $\mathcal{C}_{\mathcal T_1}$ if
  $\|\mvp_1,\mvp_2,\mvp_3,\mvp\| > 0$.
\item $\mvp$ is on $\mathcal{C}_{\mathcal T_1}$ if
  $\|\mvp_1,\mvp_2,\mvp_3,\mvp\| = 0$.
\end{itemize}

There are exactly two antipodal points that are equidistant to $p_1$,
$\mvp_2$ and $\mvp_3$. We define the spherical circumcenter of ${\mathcal
  T}_1$ as  the point $\mvc_{\mathcal T_1}$  that
is equidistant to $\mvp_1$, $\mvp_2$ and $\mvp_3$:
$$\dd(\mvp_1,\mvc_{\mathcal T_1}) = \dd (\mvp_2,\mvc_{\mathcal T_1}) = \dd (\mvp_3,\mvc_{\mathcal T_1})$$
and that is inside $\mathcal{C}_{\mathcal T_1}$. This correspond to one
of the two antipodal points that is the closets to $\mvp_1$, $\mvp_2$ and $\mvp_3$.

Consider a point set $\pointSet= \{\mvp_1,\dots,\mvp_n\}$ of $n$ points of $S$.
A triangulation $T(\pointSet)$ of $\pointSet$ is a set
of $2n-4$ non overlapping spherical triangles 
$$T(\pointSet) = \{ {\mathcal T}_1, {\mathcal T}_2,\dots, {\mathcal T}_{2n-4}\}$$
 that exactly covers $S$ with all points of $\pointSet$ being among the vertices
of the triangulation.  
 
A spherical triangle $\mathcal T_j$ is Delaunay if its circumcircle is empty i.e. if
no point $\mvp_i$ of $\pointSet$ lies inside $\mathcal T_j$. 
The Delaunay triangulation $\DT(\pointSet)$ is such
that every triangle $\mathcal T_j$  of $DT(\pointSet)$ is
Delaunay. This construction is an actual Delaunay triangulation \cite{renka1997algorithm}. 
An interesting interpretation of this kernel starts with the 3D
orientation predicate that consist in computing the sign
of the volume of tetrahedron formed by points 
${\bf p}_j(x_i,y_i,z_i)$, $j=1,\dots,4$:
\begin{equation}
\label{eq:ori3D}
\mbox{sign}
\left|
\begin{array}{cccc}
1 &1 &1 &1\\
x_1 &x_2 &x_3 &x_4\\
y_1 &y_2 &y_3 &y_4\\
z_1 & z_2 & z_3 & z_4
\end{array}
\right|
\end{equation}
The 2D 'in-circle' predicate that tells if point $\mvp_4$ belongs to the
circum-circle of triangle formed by points $\mvp_1$, $\mvp_2$,
$\mvp_3$ can be written as
\begin{equation}
\label{eq:circ2D}
\mbox{sign}
\left|
\begin{array}{rrrr}
1 &1 &1 &1\\
x_1 &x_2 &x_3 &x_4\\
y_1 &y_2 &y_3 &y_4\\
x_1^2+y_1^2 & x_2^2+y_2^2 & x_3^2+y_3^2  & x_4^2+y_4^2 
\end{array}
\right|
\end{equation}
Predicate \eqref{eq:circ2D} has a form that is close to the one
of \eqref{eq:ori3D}. This is an expression of the standard link between 3D convex
hulls and 2D Delaunay triangulations: assume a 2D 
triangulation and lift it to the paraboloid $z = x^2+y^2$. Then a
2D triangle is Delaunay if it belong to the convex hull of the lifted
triangulation. In other words, a point $\mvp(x,y)$ belongs to the circumcircle of
a triangle $t(\mvp_1,\mvp_2,\mvp_3)$ if its lifting $\mvp'(x,y,x^2+y^2)$ on the paraboloid is
below the plane defined by the lifted triangle
$t'(\mvp'_1,\mvp'_2,\mvp'_3)$. This is verified by computing the
sign of the volume of tetrahedron with vertices
$\mvp'_1,\mvp'_2,\mvp'_3,\mvp'_4$ using Equation \eqref{eq:ori3D}. In the case of a
triangulation on a unit sphere, predicate \eqref{eq:ori3D}  becomes
\begin{equation}
\label{eq:ori3Ds}
\mbox{sign}
\left|
\begin{array}{cccc}
1 &1 &1 &1\\
x_1 &x_2 &x_3 &x_4\\
y_1 &y_2 &y_3 &y_4\\
z_1 & z_2 & z_3 & z_4
\end{array}
\right| = 
\mbox{sign}
\left|
\begin{array}{rrrr}
1 &1 &1 &1\\
x_1 &x_2 &x_3 &x_4\\
y_1 &y_2 &y_3 &y_4\\
\sqrt{1-x_1^2-y_1^2} & \sqrt{1-x_2^2-y_2^2} & \sqrt{1-x_3^2-y_3^2} & \sqrt{1-x_4^2-y_4^2}
\end{array}
\right|.
\end{equation}
The lifting here is on the sphere and not on the paraboloid and the
construction that is proposed is a Delaunay triangulation. 

\section{A parallel Delaunay Kernel}
\label{sec:ker}
A triangulation $T(\pointSet)$ of $\pointSet$ is a set
of non overlapping triangles that exactly covers the convex hull
$\Omega(\pointSet)$ with all points of $\pointSet$ being among the vertices
of the triangulation.  

Delaunay triangulations are
popular in the meshing community because fast algorithms exist that allow to
generate $\DT(\pointSet)$ in ${\mathcal O(n\log{n}})$ complexity.
\label{sec:ker}
\begin{figure}[h!]
 \begin{center}
 \begin{tabular}{ccc}
 \scalebox{0.5}{\includegraphics{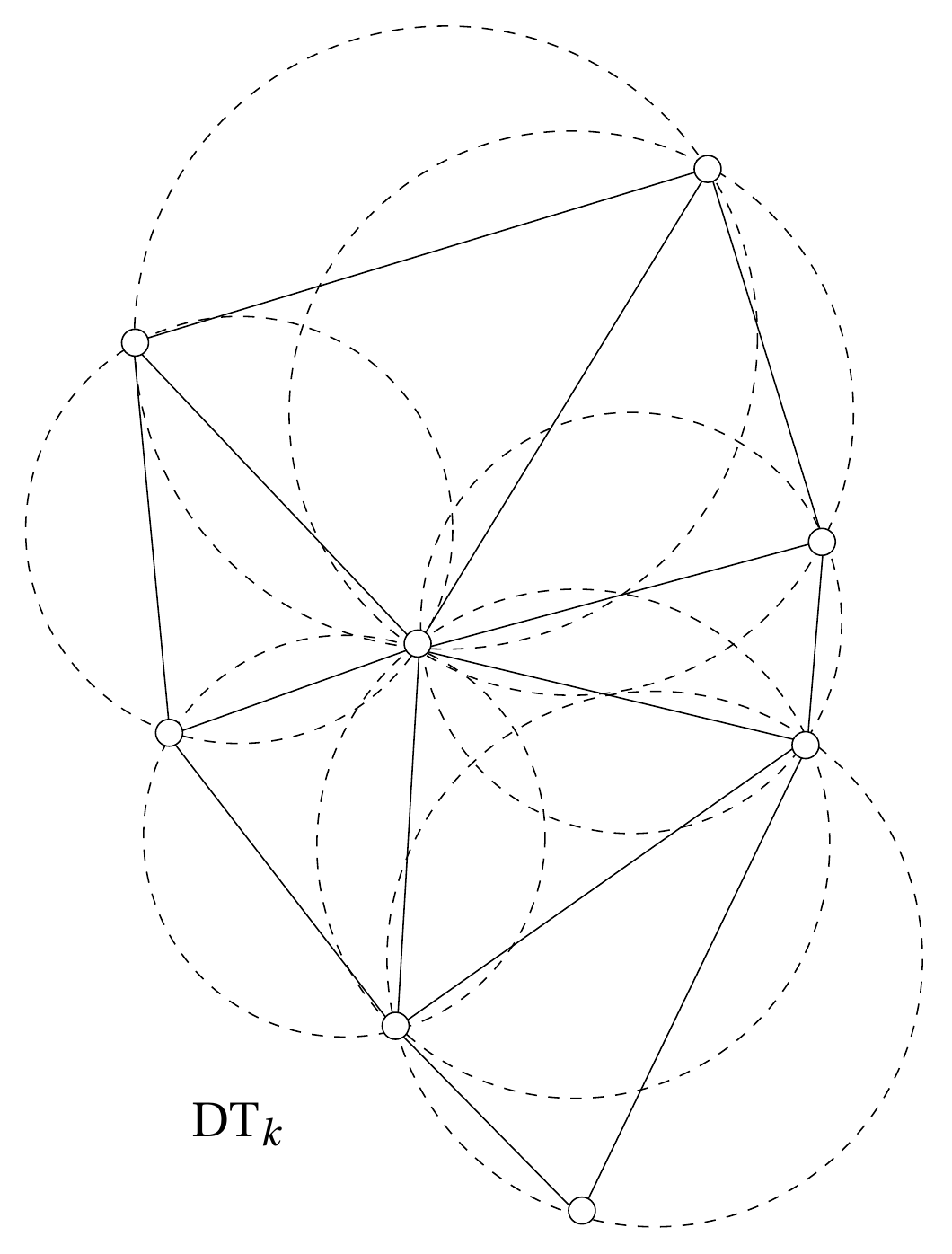}}&
 \scalebox{0.5}{\includegraphics{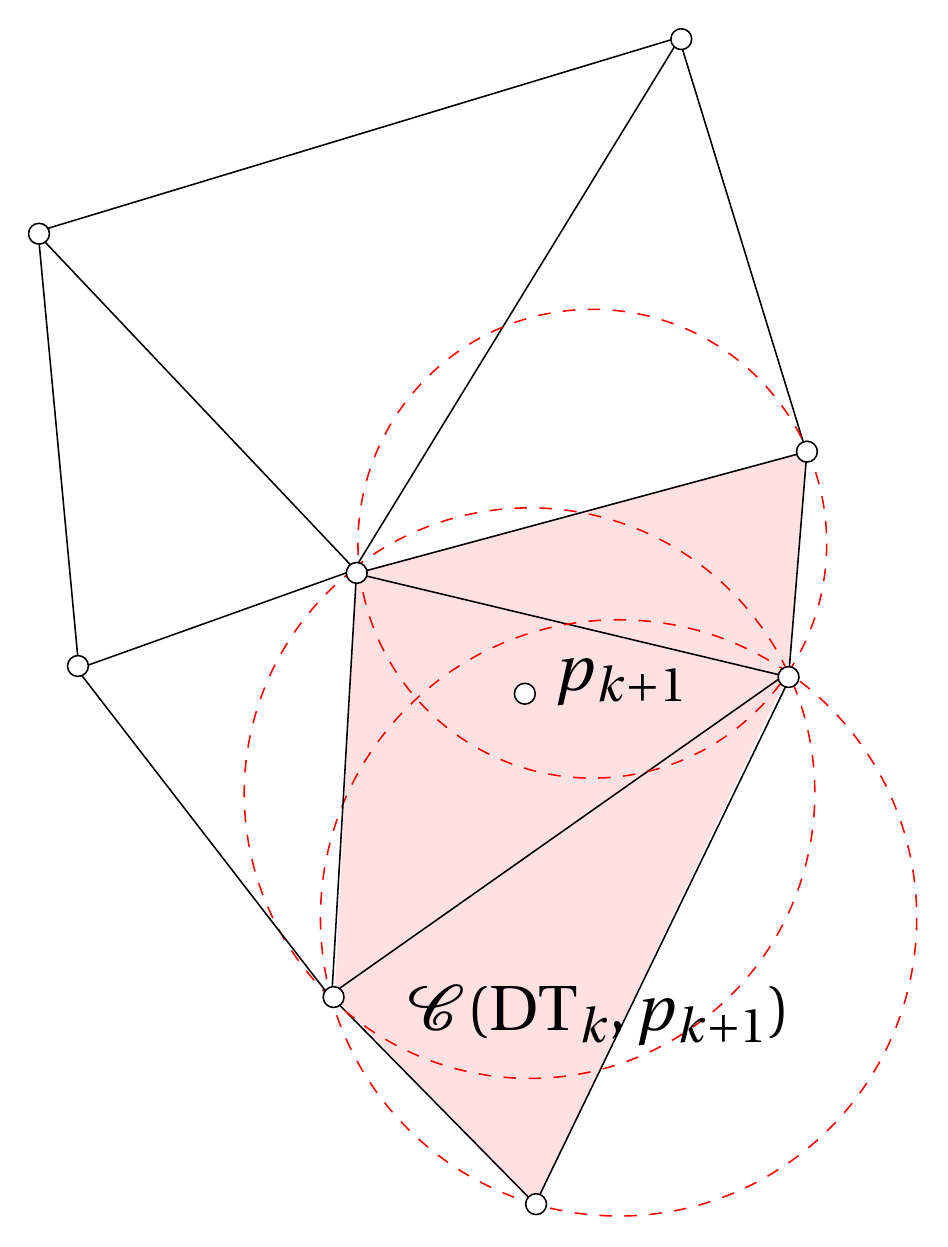}}&
 \scalebox{0.5}{\includegraphics{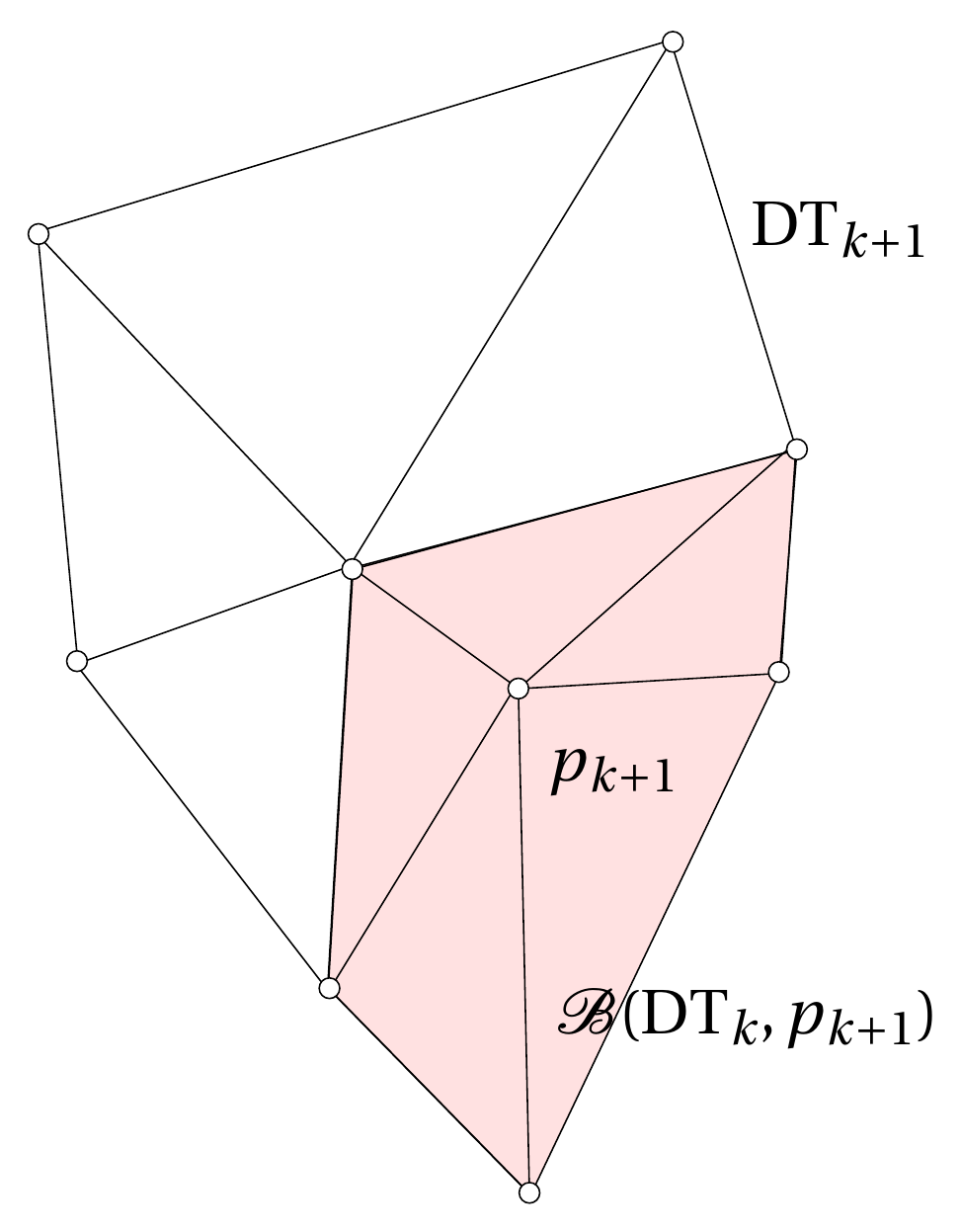}}
\end{tabular}
\caption{Delaunay triangulation ${\DT}_k$ (left), Delaunay cavity
  ${\mathcal C}_p({\DT}_k,{\mvp}_{k+1})$ (center) and
  ${\DT}_{k+1} = {\DT}_{k} - {\mathcal C}({\DT}_k,{\mvp}_{k+1}) + {\mathcal
    B}(\DT_k,\mvp_{k+1})$ (right).\label{fig:cavity}}
 \end{center}
\end{figure}

Let ${\DT}_k$ be the Delaunay
triangulation of a point set
$\pointSet_k = \{\mvp_1,\dots,\mvp_k\} \subset \reals^d$.  The \emph{Delaunay
  kernel} is a procedure that allow the incremental insertion of a given
point ${\mvp}_{k+1} \in \Omega(\pointSet_k)$ into ${\DT}_k$ and to build the
Delaunay triangulation $\DT_{k+1}$ of
$\pointSet_{k+1} = \{\mvp_1,\dots,\mvp_k,\mvp_{k+1}\}.$ The \emph{Delaunay kernel}
can be written in the following abstract manner:
\begin{equation}
  {\DT}_{k+1} = {\DT}_{k} - {\mathcal C}({\DT}_k,{\mvp}_{k+1}) +
  {\mathcal B}(\DT_k,\mvp_{k+1}) ,\label{eq:delker}
\end{equation}
where the Delaunay cavity ${\mathcal C}({\DT}_k,{\mvp}_{k+1}) $ is the set of
all triangles whose circumcircles contain the new point $\mvp_{k+1}$ (see
Figure \ref{fig:cavity}; the triangles of the cavity cannot belong to
${\DT}_{k+1}$) and the Delaunay ball ${\mathcal B}(\DT_k,\mvp_{k+1})$ is a set
of triangles that fill the polyhedral hole that has been left empty while
removing the Delaunay cavity ${\mathcal C}({\DT}_k,{\mvp}_{k+1})$ from
$\DT_{k}$.

It is possible to build  ${\DT}_{k+1}$ in ${\mathcal O} (n \log n)$
operations. The critical operation in the Delaunay kernel procedure is
the construction of the cavity $ {\mathcal C}({\DT}_k,{\mvp}_{k+1})$. A
first element of the cavity $\tau$ is searched in $\DT_k$ and the cavity is
constructed using a depth first search algorithm that starts at
$\tau$. Sorting the point set $\pointSet$  in such a
way that two successive points in the set are close to each other
geometrically allows to find $\tau$ in a number of operations that
actually does not depend on $n$. We use her the standard BRIO sort
that consist in sorting increasingly large subsets of $\pointSet$
along Hilbert curves. The points are not sorted all at once in order
not to produce large cavities.

Building the cavity actually takes a
number of operations that essentially depends on the dimension (2D or
3D). So, after the point set is sorted (${\mathcal O}(n\log n)$
operations), the insertion of the points takes ${\mathcal O}(n)$ operations. 

It is possible to construct a multi-threaded version of the Delaunay
kernel \cite{remacle2015two}. Assume $M$ computational threads that aim at inserting $M$ points in the
triangulation at the same time. At the end, each thread is going to insert
${n\over M}$ points.  The situation is of course not that simple: two points $\mvp_i$ and $\mvp_j$
can only be inserted at the same time in $\DT_k$ if their respective
Delaunay cavities ${\mathcal C}({\DT}_k,{\mvp}_{i})$ and
${\mathcal C}({\DT}_k,{\mvp}_{j})$ do not overlap, i.e., if they do not have
triangles in common.
A non-overlapping situation is more likely to happen if points $\mvp_i$ and
$\mvp_j$ are not close geometrically. For that purpose, we split the Hilbert
curve into $M$ equal parts and assign each part to one thread. 

The \emph{multithreaded Delaunay kernel} can be written in the following
abstract manner:
\begin{equation} 
{\DT}_{k+1} = {\DT}_{k} + \sum_{i=0}^{M-1} \left[{-\mathcal
      C}({\DT}_k,{\mvp}_{k + i{n \over M}}) + {\mathcal B}(\DT_k,\mvp_{k + i{n
        \over M}})\right].\label{eq:delker}
\end{equation}

We have implemented the multithreaded Delaunay kernel using OpenMP
\cite{openmp}. Two OpenMP barriers were used at each iteration $k$. A first
barrier is used after the computation of the $M$ cavities: every thread $i$
has to complete its cavity ${\mathcal C}({\DT}_k,{\mvp}_{k + i{n \over M}})$ at
iteration $k$ in order to be able to verify that the cavity does not overlap
other cavities. When several cavities overlap, only the point corresponding
to the smallest thread number is processed. The other points are delayed to
the next iteration.  A second barrier is used after the construction of
$ {\mathcal B}(\DT_k,\mvp_{k + i{n \over M}})$: every thread has to finish
computing the Delaunay kernel in order to start iteration $k+1$ with a valid
mesh. This simple procedure usually produces speedups of $3$ on a
quad-core computer. A two-level version of that procedure is now
available in the 3D Delaunay mesher of Gmsh \cite{geuzaine2009gmsh}.


\section{Building a valid 1D mesh}
\begin{figure}
  \begin{center}
  \begin{tabular}{cc}
    \setlength{\fboxsep}{0pt} \fbox{\includegraphics[width=7cm]{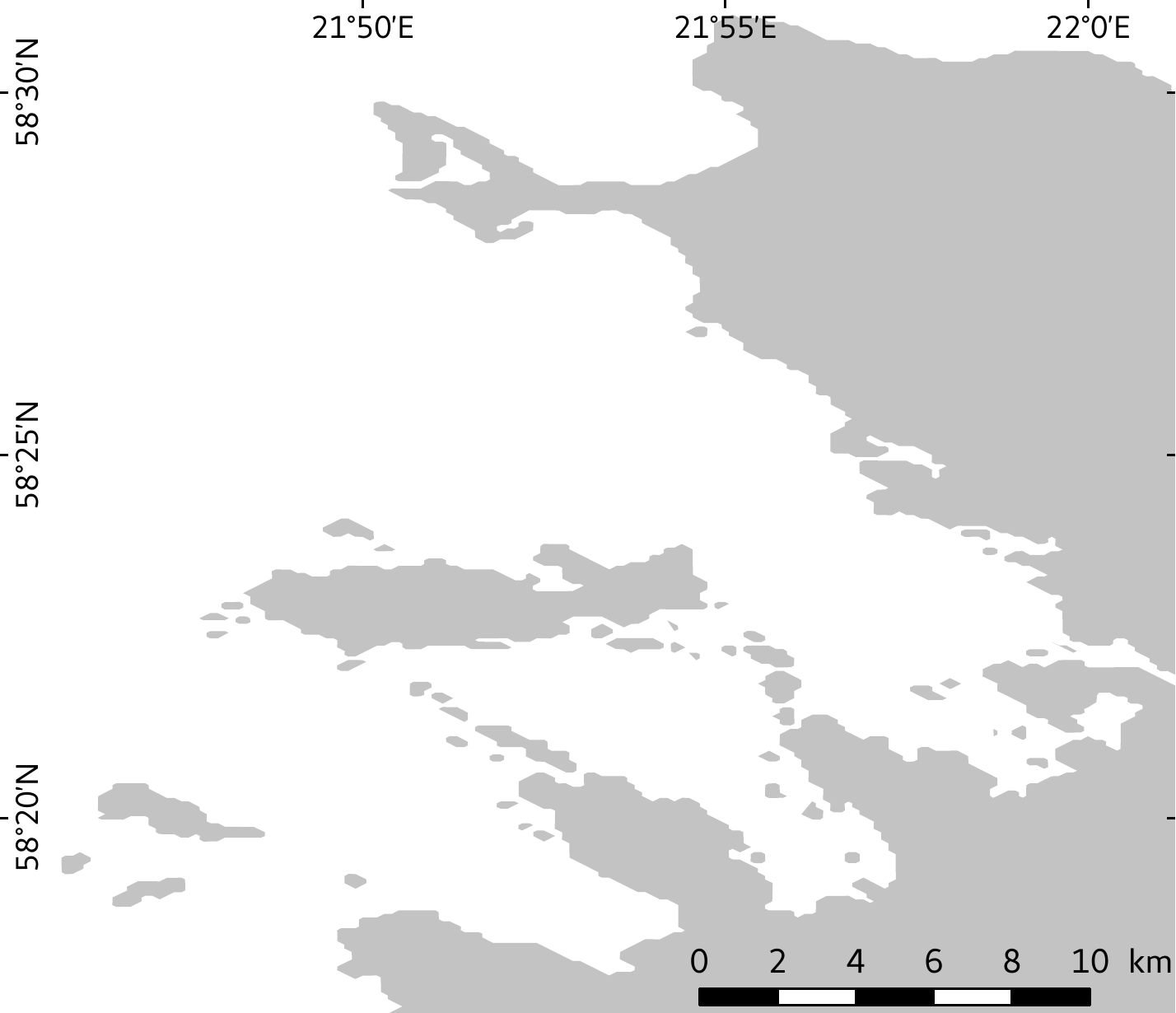}}&
    \setlength{\fboxsep}{0pt} \fbox{\includegraphics[width=7cm]{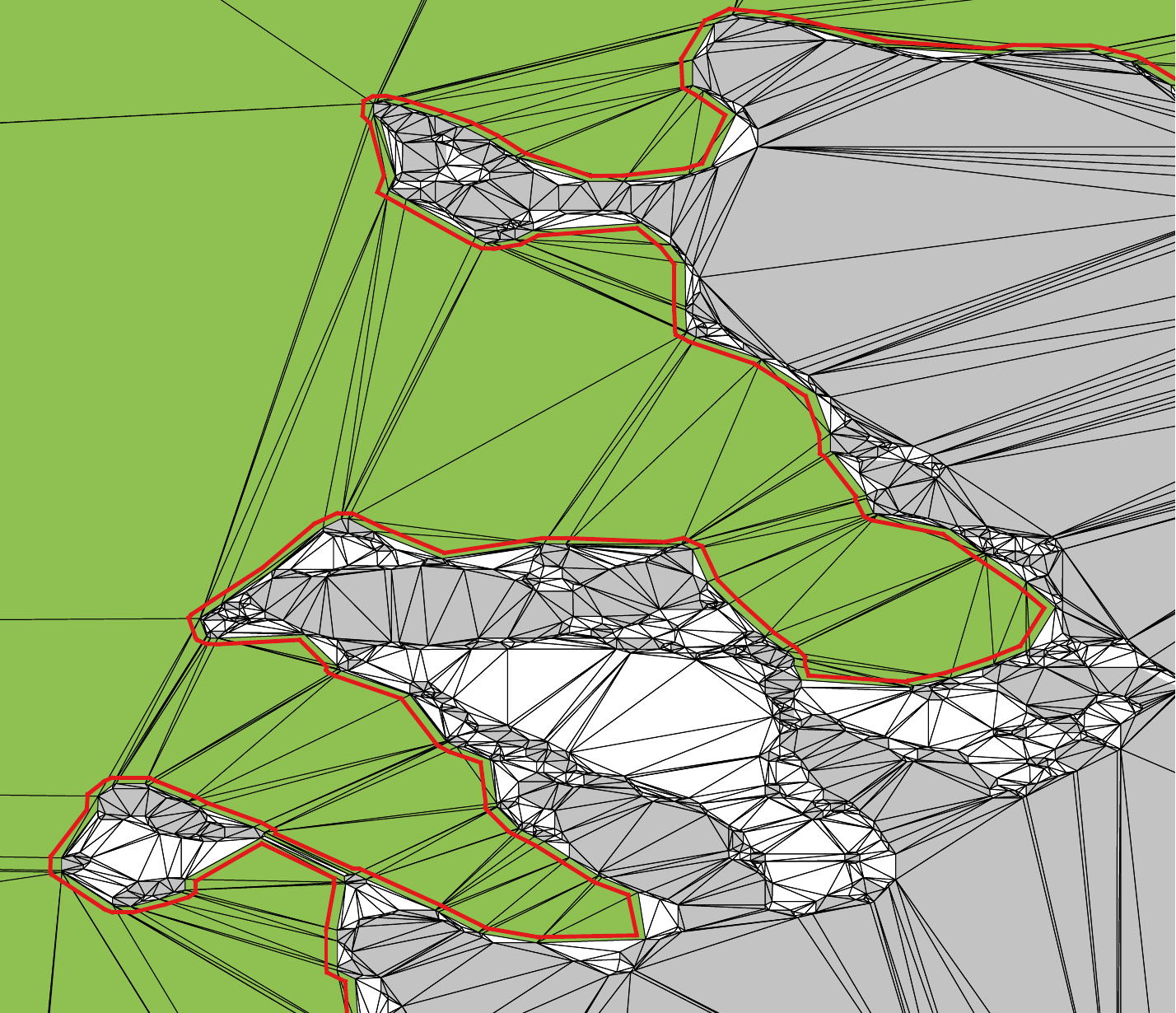}}\\
    \footnotesize{(a) high-resolution raw data} &
    \footnotesize{(b) and (c) generation of the coarse (1500m) geometry} \\[0.3cm]
    \setlength{\fboxsep}{0pt} \fbox{\includegraphics[width=7cm]{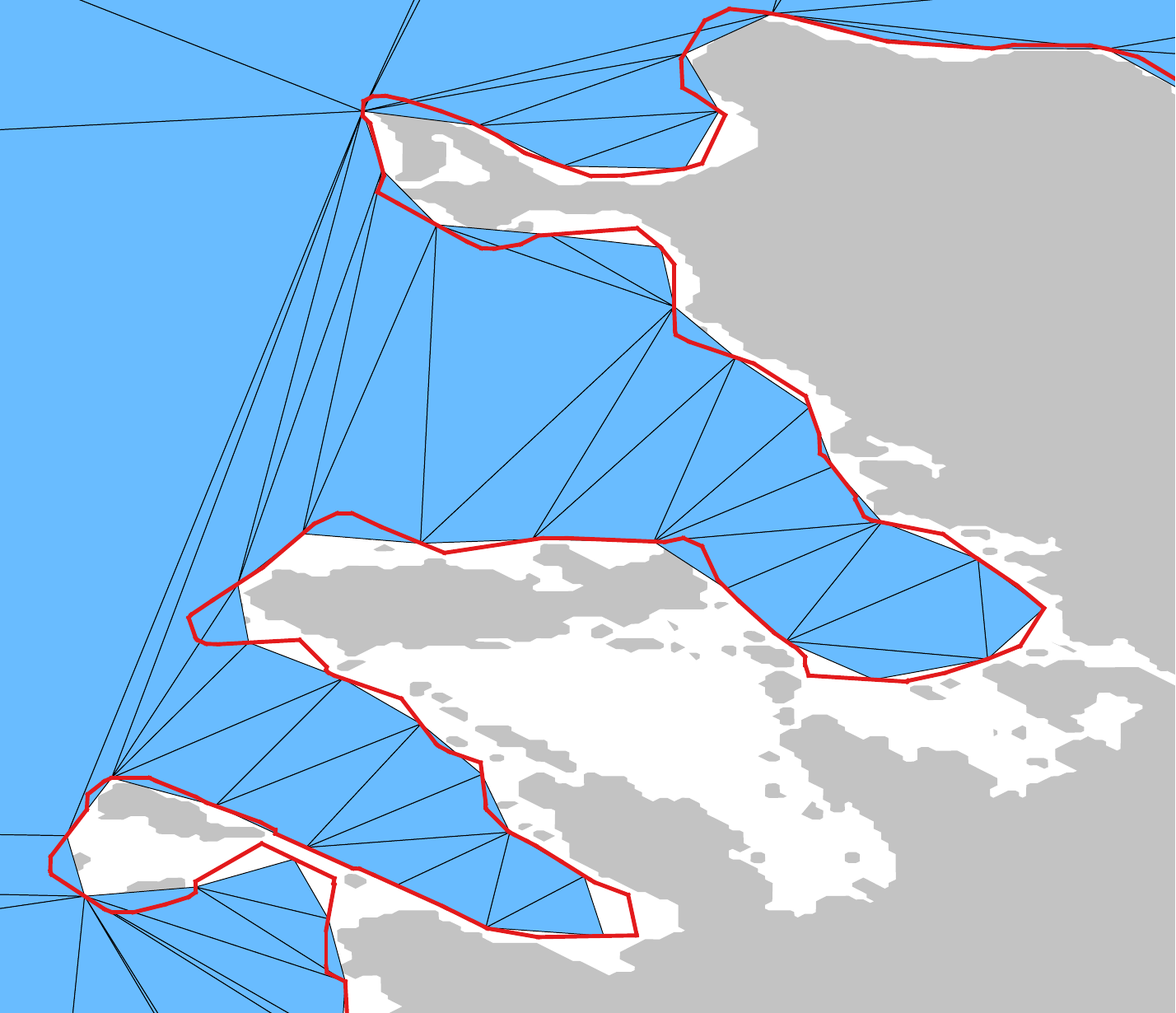}}&
    \setlength{\fboxsep}{0pt} \fbox{\includegraphics[width=7cm]{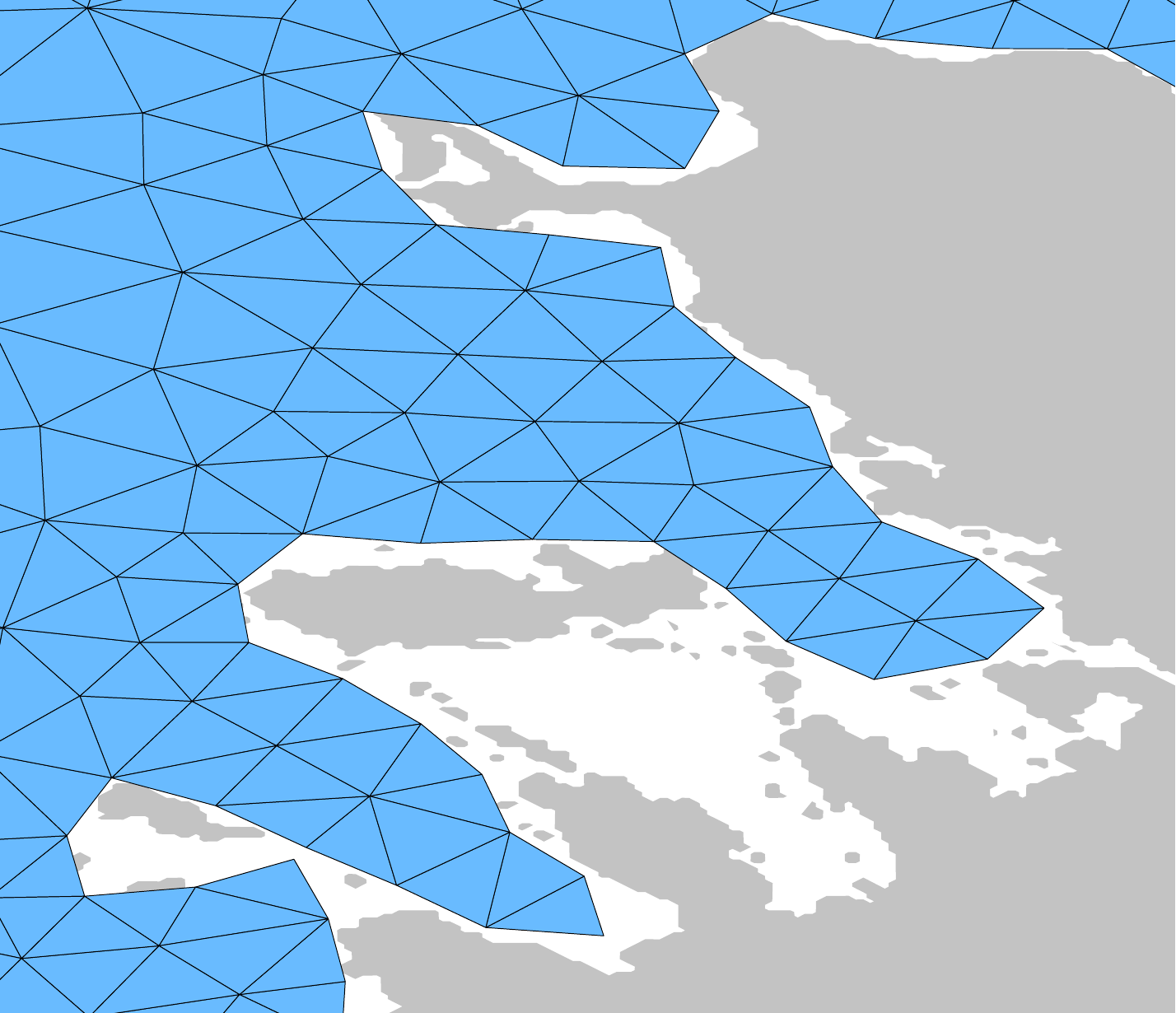}}\\
    \footnotesize{(d) discretization of the coarse geometry} &
    \footnotesize{(e) final mesh for a resolution of 1500m} \\[0.3cm]
    \hline\\
    \setlength{\fboxsep}{0pt} \fbox{\includegraphics[width=7cm]{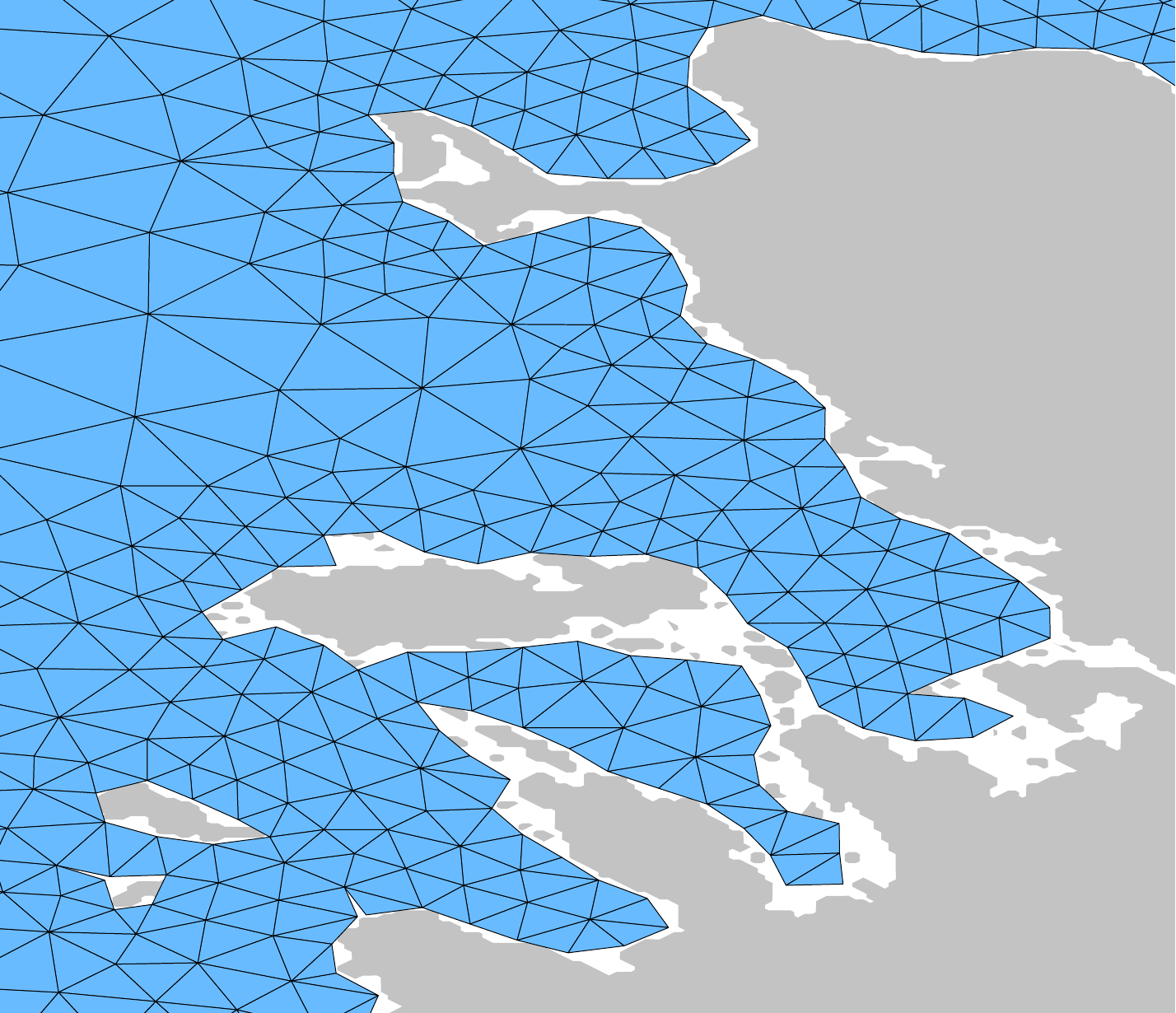}}&
    \setlength{\fboxsep}{0pt} \fbox{\includegraphics[width=7cm]{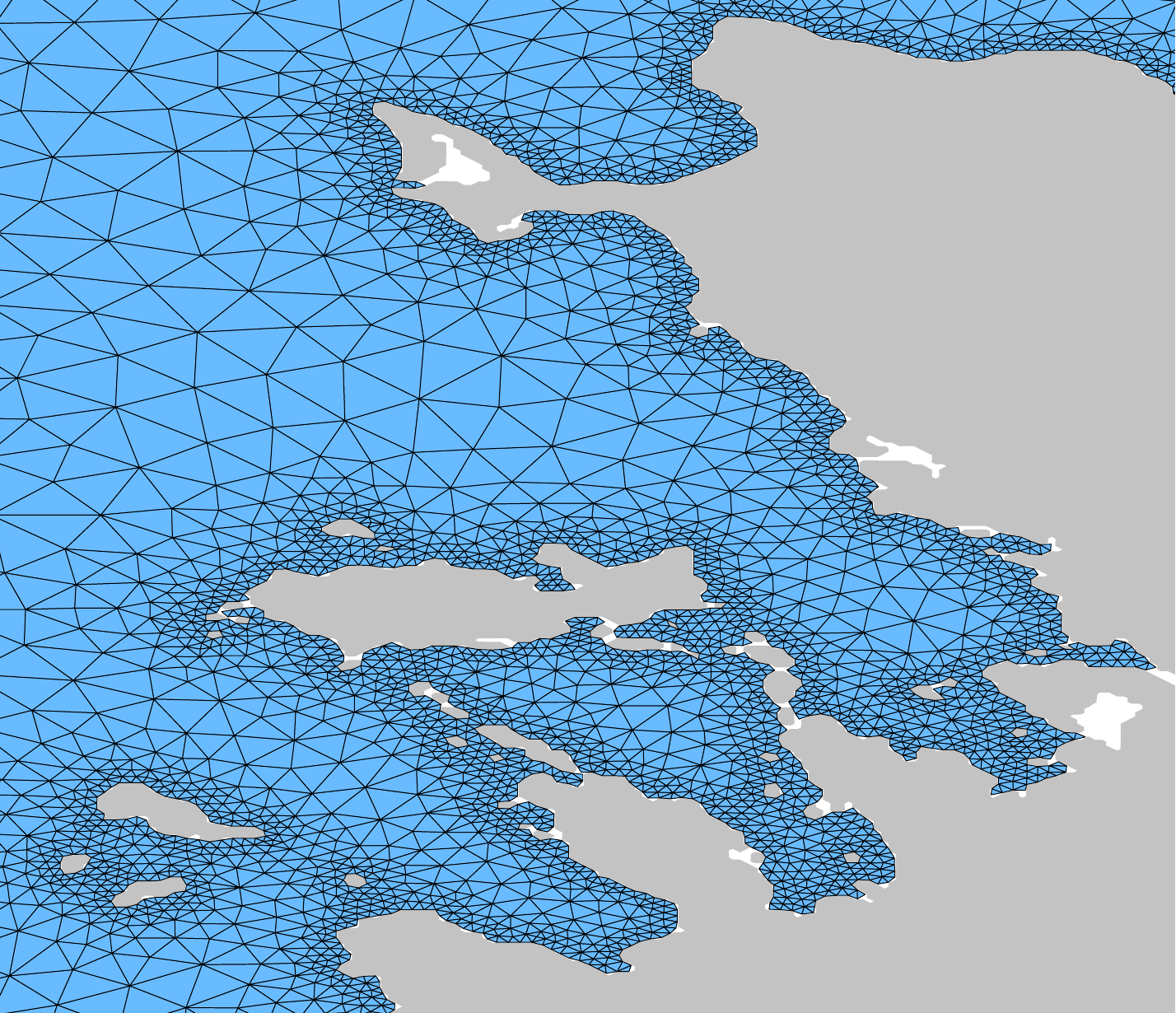}}\\
    \footnotesize{final mesh for a resolution of 750m along the coastlines} &
    \footnotesize{final mesh for a resolution of 150m along the coastlines}
  \end{tabular}
\caption{Generation of a mesh coarser than the initial data\label{fig:unrefine}.}
  \end{center}
\end{figure}
The domain boundaries are extracted from geographical coastline databases.
Due to their fractal nature, coastal shapes can be extremely complex and contain many small details.
A major difficulty when dealing with coastline is that the resolution of the geographical database doest not match the desired mesh resolution.
For example, the freely-available GSHHG \cite{wessel2013gshhg} database has a maximal resolution bellow $100$m which is much too small for many numerical applications.
In ocean modeling, a good practice would be that there exists no isolated island and no channel in the domain that are smaller than the mesh size.
Small features should be removed from the description of the domain,
eventually in an automatic fashion. If such small features exist, (i)
badly shaped elements would be created in such channels or around
those islands and (ii) the model will not be able anyway to capture relevant physics on features that are below mesh size. 

Assume a mesh size field $h(x,y,z)$ that defines local sizes of mesh
edges at point $x,y,z$ of the domain.
We propose here a procedure that automatically creates a valid 1D mesh
that: (i) does not overlap itself, (ii) contains no edges
significantly smaller than $h$ and (iii) that defines the contour of a domain without water channels that are
significantly narrower than $h$ (and in consequence, the boundary of
the domain does no contain any angles that are strongly acute).

Figure \ref{fig:unrefine} illustrates the different steps involved to build a coarse mesh from the high-resolution raw GSHHG data of the western coastline of the Saaremaa Island (Estonia) in the Baltic sea.
\begin{enumerate}
  \item[(a)]
    A high-resolution raw dataset is initially loaded. A mesh size
    field $h(x,y,z)$ is defined by the user. Any edge of the coastline
    data that is smaller than $h$ is refined.
  \item[(b)]
    A Delaunay mesh is built on the sphere with all the boundary
    points of the raw dataset.  The user then selects one point $p$
    that belongs to the region that has to be meshed. 
    The unique curvilinear triangle $\tau$ that contains $p$ is then found by walking
    into the triangulation \cite{devillers2002walking} and  a depth-first search algorithm is
    applied to traverse the mesh, starting at root $\tau$. The
    algorithm stops whenever an edge smaller than $h$ is crossed
    (green triangles on sub-Figure (b) of Figure
    \ref{fig:unrefine}). This procedure defines a new set of edges
    that form a coarsened version of the domain that has no small
    features. Note here that the initial boundary of the domain cannot
    be traversed because edge size of the initial boundary is
    guaranteed to be smaller than $h$ (see point (a) above). 
    Finally, note that the Delaunay mesh guarantees that
   two points closer to each other than $h$ in the initial triangulation
   will either be connected by an edge of the triangulation or connected by a series of edges smaller than $h$.  This property prevents us to walk across narrow channels or across the original boundaries. 
  \item[(c)]
    The external boundary of the green triangulation defines the new boundary of
    the domain of interest where all small features have been removed.
    At this point, the remaining isolated islands smaller than the prescribed mesh size are removed.
    There may exist edges inside the domain that are smaller than $h$.
    Such small edges are always bounded by two green triangles: they
    are embedded in the domain. To avoid this situation, which is
    problematic for the following steps, the whole boundary line (in
    red) is slightly (e.g. by $h/10$) 
    shifted inside the domain.
  \item[(d)]
    The red line is the actual boundary of the domain. It is
    discretized with a prescribed mesh size. 
    A Delaunay mesh is built with those points and the boundary edges are recovered by performing swaps.
  \item[(e)]
    Finally, a Delaunay refinement algorithm described below is applied to generate the points inside the domain.
\end{enumerate}
A side benefit of this approach is that it requires only a mesh size
field $h(x,y,z)$, the coordinates of one point 
inside the domain, and a series of (close) boundary points.
In particular the initial boundary lines do not have to describe a correct topology.
In other words, it is not a problem if two initial coastlines intersect each other.
This makes it easy to mix various datasets or cut through a domain.

\section{Multi-threaded Delaunay refinement}
At that point, a valid 1D mesh has been produced. The following stage
of the meshing algorithm consist in generating the surface mesh. For
that, an ``empty mesh'' that contains all vertices of the 1D mesh is
constructed on the sphere. Then, every edge of the 1D mesh that is not
present in the empty mesh are recovered using edge swaps. 

A Delaunay refinement procedure is applied to generate internal
points of the domain. Here, we use an edge based approach. Every
internal edge of the domain is ``saturated'' in the following way. 
Consider an edge ${\bf pq}$ of length $L = \|{\bf q}-{\bf p}\|$. 
A point ${\bf x}$ of ${\bf pq}$ has the form
$${\bf x}(u) = {\bf p}(1-u) + {\bf q} u~~,~~u \in [0,1].$$
The adimensional quantity 
$$\rho(u) = {L \over h({\bf x}(u))}$$
represents the mesh density at point ${\bf x}$ i.e. the number of
points per unit of length that should be used to saturate an edge at
point ${\bf x}$.
The following primitive
$$
\delta(t) = \int_0^t \rho(u) du 
$$
is very useful in what follows. We first note that $\delta(1)$ is the
adimensional length of ${\bf pq}$. Its rounded up
value $N = \ceil{\delta(1)}$ represents
the number of subdivisions that is required to saturate
edge ${\bf pq}$ i.e. to split ${\bf pq}$ with $N$ sub-segments that have an
adimensional length close to one. 

An adaptative trapezoidal integration scheme is used to
compute $\delta(t)$ with a prescribed accuracy. 

\label{sec:delnlogn}
\begin{algorithm}
     \SetKwData{Left}{left}\SetKwData{This}{this}\SetKwData{Up}{up}
     \SetKwFunction{Union}{Union}\SetKwFunction{FindCompress}{FindCompress}
     \SetKwInOut{Input}{input}\SetKwInOut{Output}{output}      
      \Input{A mesh size field function $h({\bf x})$, an edge ${\bf
        pq}$ and a prescribed accuracy $\epsilon$.}
      \Output{A discrete representation of $\delta(t) = \int_0^t
        \rho(u) du$ as a list of $K$ points $(t_j,\delta_j)$, $j=1,\dots,K$.}
      \BlankLine
      Initialize a stack with one tuple $\{0,1,h({\bf x}(0)),h({\bf
      x}(1))\}$ with ${\bf x}(u) = {\bf p}(1-u) + {\bf q}u.$\;

      Initialize an empty vector of tuples that contains the piecewise
      linear representation of the size field\;
      \BlankLine
      \While{stack is not empty} {
         take tuple $\{ u_1,u_2,h_1,h_2 \}$ at the top of the stack \;
         pop this tuple out of the stack \;
         compute $u_{12} = (u_1+u_2)/2$ and $h_{12} = h ( {\bf x} (u_{12}))$\; 
         
         \If{ $\left| 1 / h_{12} - 2/ (h_1+h_2)\right|  > \epsilon  \|{\bf p} - {\bf q}\| $ }{
             push tuple $\{u_{12}, u_{2}, h_{12},h_{2}\}$ at the top of
             the stack\;

             push tuple $\{t_1, t_{12}, h_1,h_{12}\}$ at the top of
             the stack\;
       }
       \Else{
           add tuple $\{ u_1,u_2,h_1,h_2 \}$ to the result vector\;
       }
     }
   
\caption{Adaptive trapezoidal rule that computes a piecewise linear
  representation of the size field along an edge ${\bf pq}$.\label{algo:trapez}}
\end{algorithm}

The primitive is numerically approximated by a piecewise linear
function (see Algorithm \ref{algo:trapez}).
The $N-1$ points that form $N$ segments of
adimensional size close to one on ${\bf pq}$ are situated at positions
${\bf x}(t_j)$, $j=1,\dots,N-1$
with $t_j$ computed in such a way that
$$\delta(t_j) = j {\delta(1) \over N}.$$
Points $t_j$ are computed using the piecewise approximation of $\delta(t)$. 

At that point, it is interesting to give some justifications for that
rather complex procedure. Mesh size functions $h({\bf x})$ can be complex
functions involving heavy computations. In the cas of ocean modeling,
$h$ may involve the distance to coastlines as well as local bathimetry
\cite{}. It is thus mandatory to minimize the number of function calls
to $h$. 

\begin{figure}[t!]
 \begin{center}
 \begin{tabular}{c}
\includegraphics[width = 0.98\linewidth]{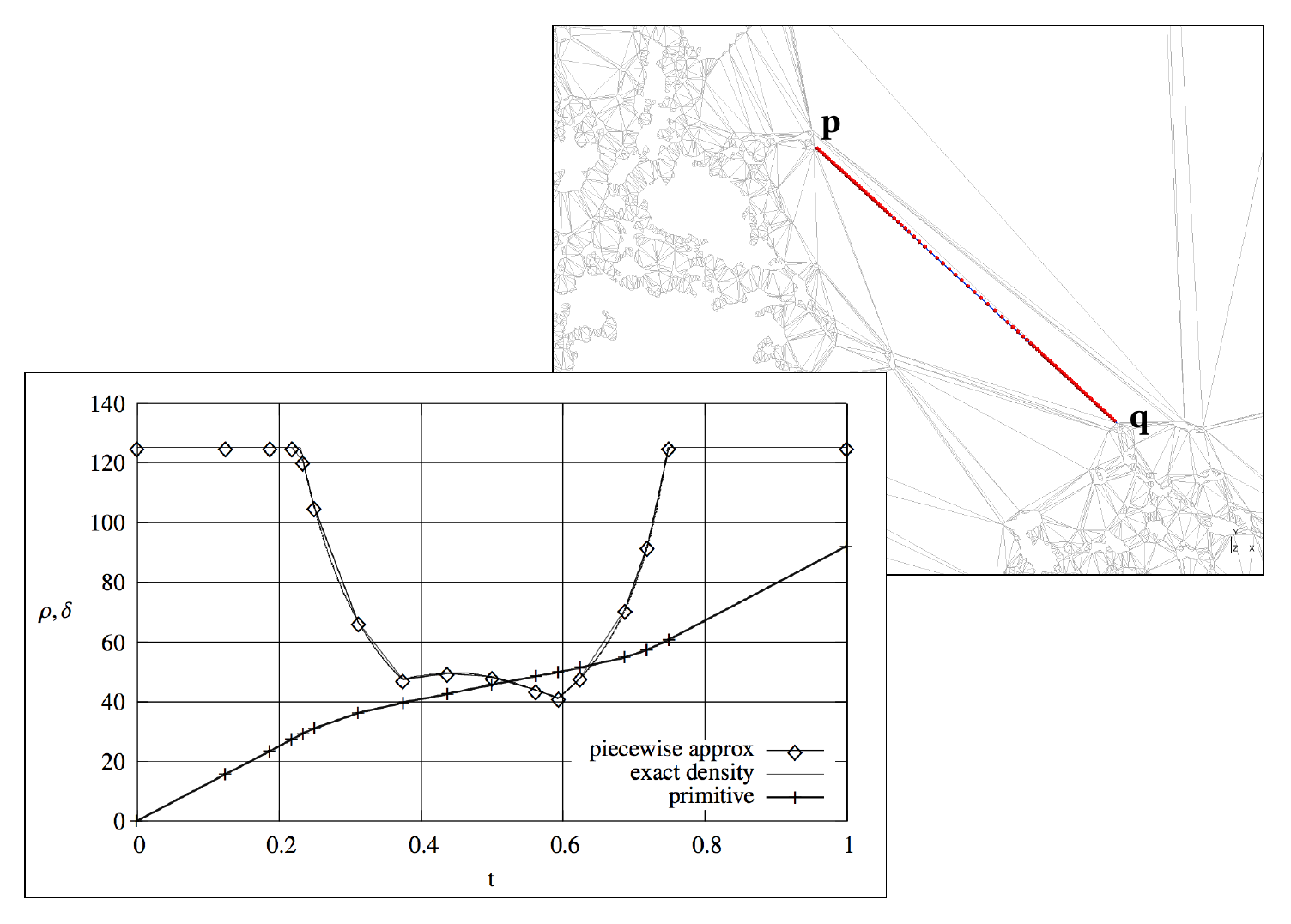}
\end{tabular}
\caption{Initial ``empty mesh'' of the baltic sea. Picture shows a
  zoom around an edge ${\bf pq}$. Points ${\bf x}_j$, $j=1,\dots,92$ 
  that saturate the edge have been drawn in red. 
  Computation of $\rho(t)$ and $\delta(t)$ on edge ${\bf pq}$
  are shown as well. Only $K=16$ evaluations of $h({\bf x})$
  were required to obtain a relative error of $5\%$ on $\rho$. 
\label{fig:integ}
\label{fig:balt1}}
 \end{center}
\end{figure}
 In order to illustrate that procedure, consider again the Baltic sea
and a mesh size field 
\begin{equation}\label{eq:size}
h({\bf x}) = \left\{ \begin{array}{lcl}
h_{\min} &~~\mbox{if}~~&d({\bf x}) < d_{\min}\\
h_{\min} + {d({\bf x})-d_{\min} \over d_{\max}-d_{\min}} (h_{\max}-h_{\min}) &~~\mbox{if}~~&d_{\min} \leq d({\bf x}) \leq d_{\max} \\
h_{\max} &~~\mbox{if}~~&d({\bf x}) > d_{\max}
\end{array}
\right.\end{equation}
where $d({\bf x})$ is the distance to coastline (wall distance). The
initial mesh (empty mesh) is depicted on Figure \ref{fig:balt1}.
On the same Figure, an edge ${\bf pq}$ is shown that has an
adimensional length of $\delta(1) = 92.12$ ($N=93$). Only $K=16$
points were necessary to represent $\rho$ with a prescribed accuracy
of $\epsilon = 5\%$. Both $\rho$ and $\delta$ are represented on
Figure \ref{fig:balt1}.  The corresponding
$92$ points ${\bf x}_j$ that saturate ${\bf pq}$  are plotted in red
on the Figure as well.

\begin{figure}[t!]
  \begin{center}
  \begin{tabular}{cc}
    \setlength{\fboxsep}{0pt} \fbox{\includegraphics[width=7cm]{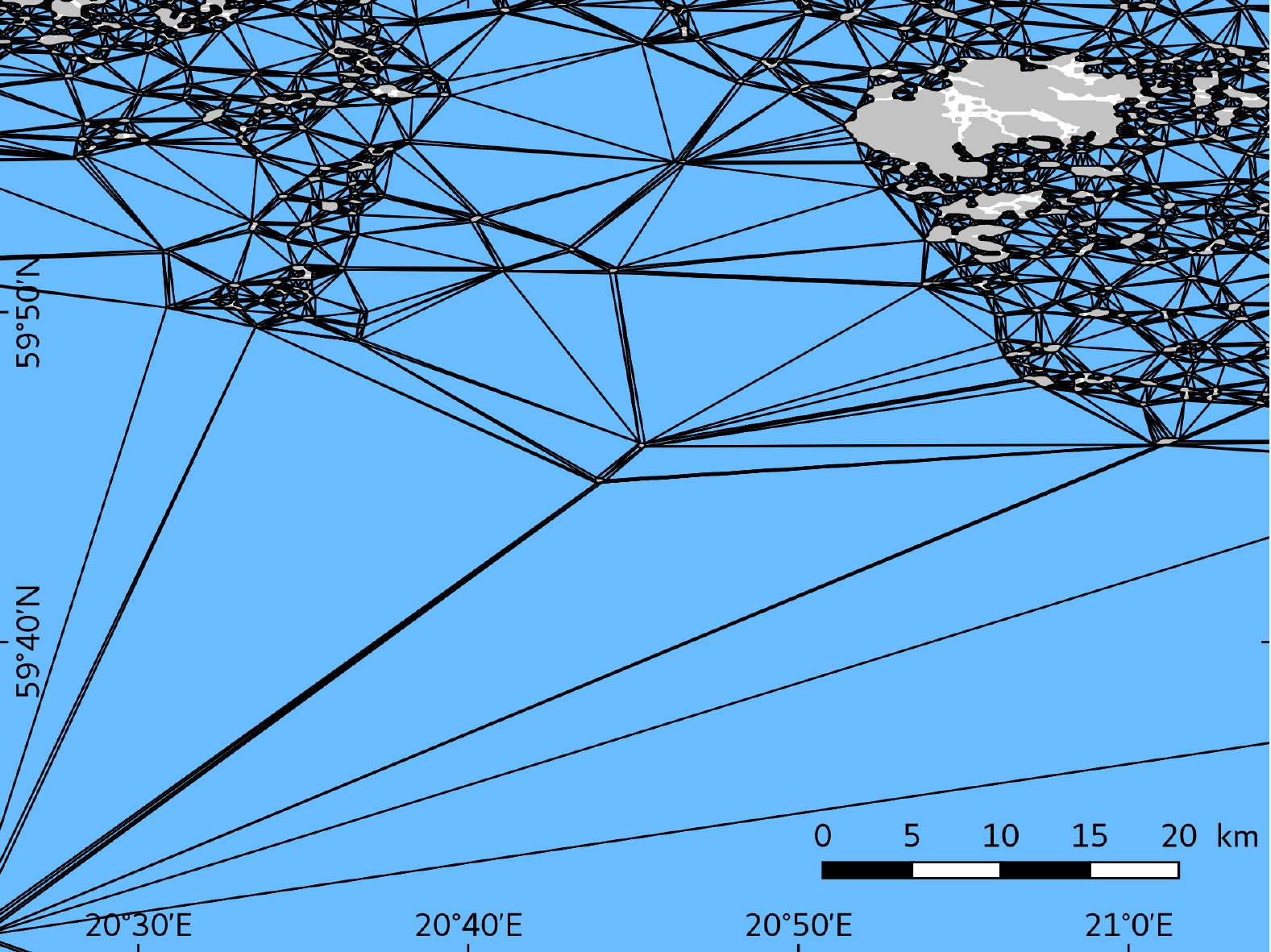}}&
    \setlength{\fboxsep}{0pt} \fbox{\includegraphics[width=7cm]{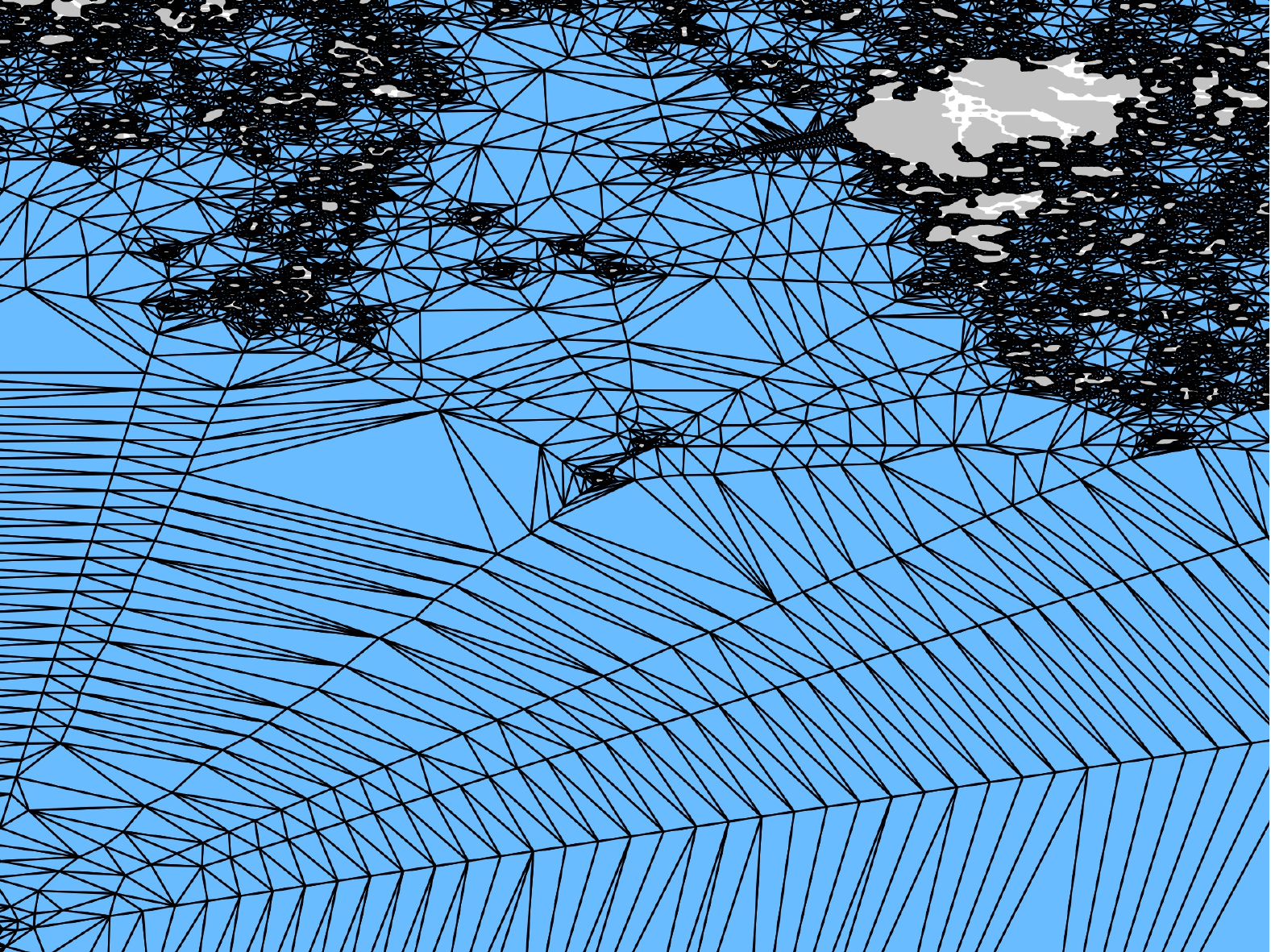}}\\
    \footnotesize{Triangulation of the 1D mesh} &
    \footnotesize{1 refinement step (471,312 points)}\\[0.3cm]
    \setlength{\fboxsep}{0pt} \fbox{\includegraphics[width=7cm]{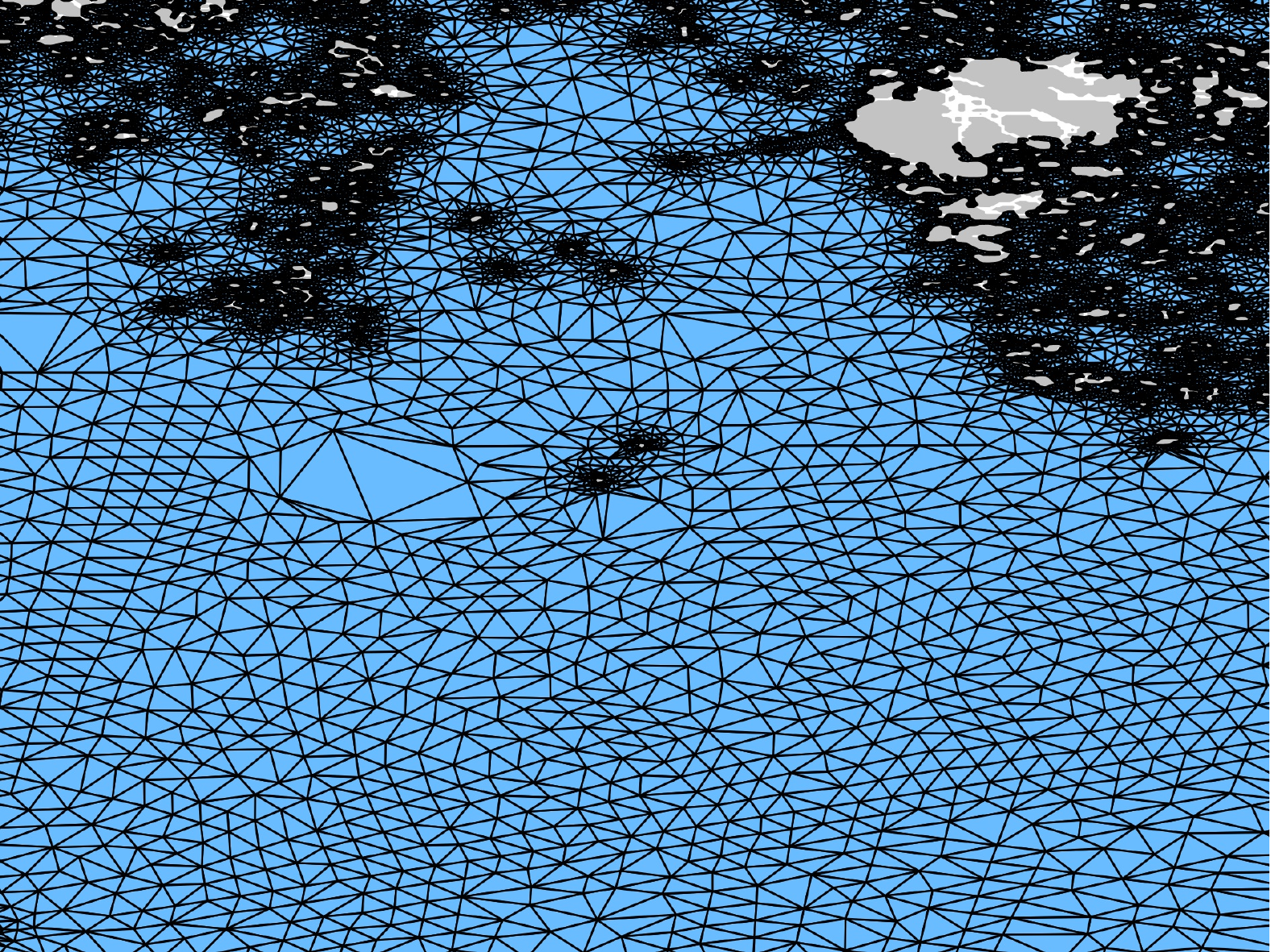}}&
    \setlength{\fboxsep}{0pt} \fbox{\includegraphics[width=7cm]{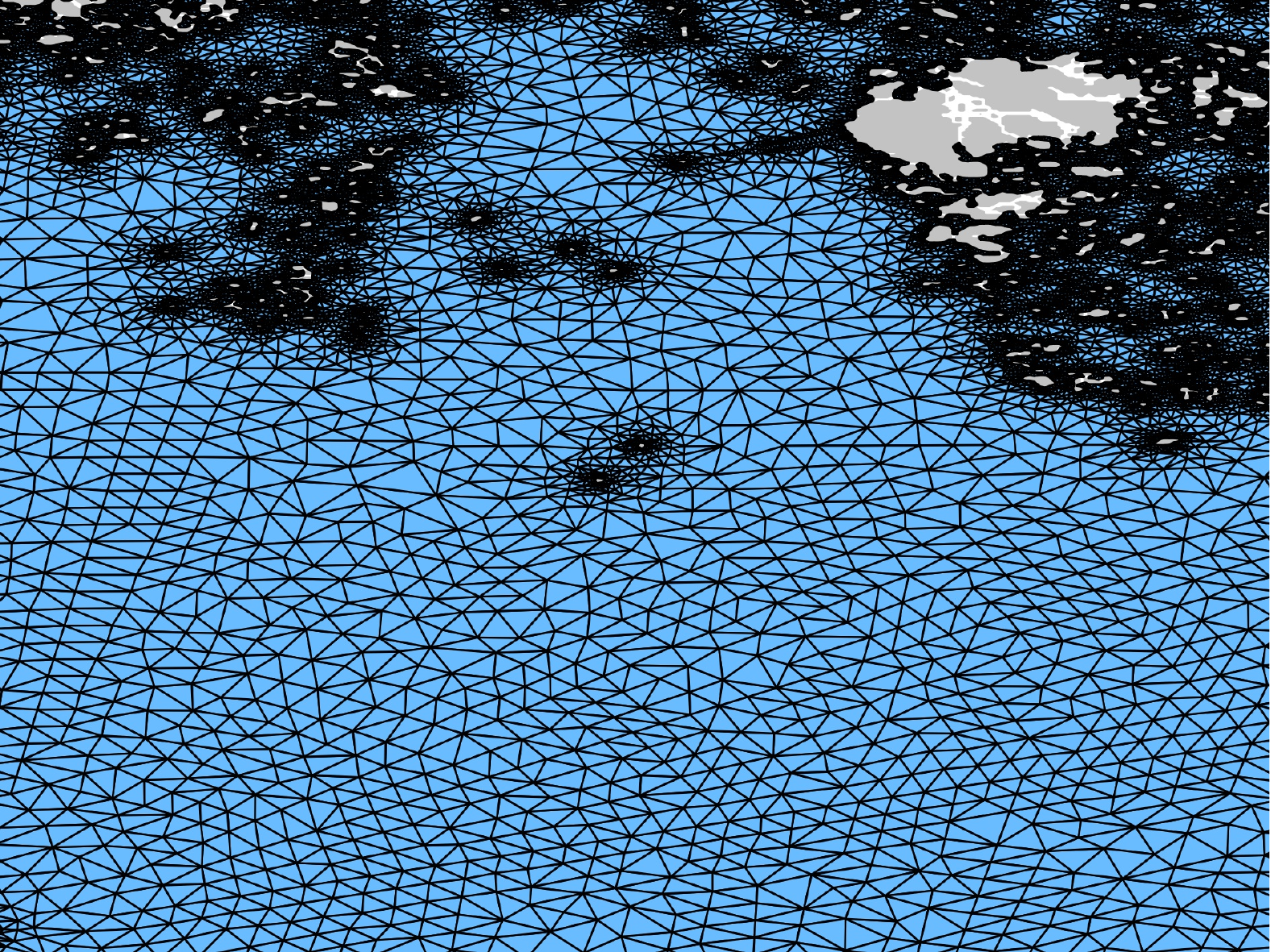}}\\
    \footnotesize{2 refinement steps (873,028 points)}&
    \footnotesize{3 refinement steps (969,283 points)}\\[0.3cm]
    \setlength{\fboxsep}{0pt} \fbox{\includegraphics[width=7cm]{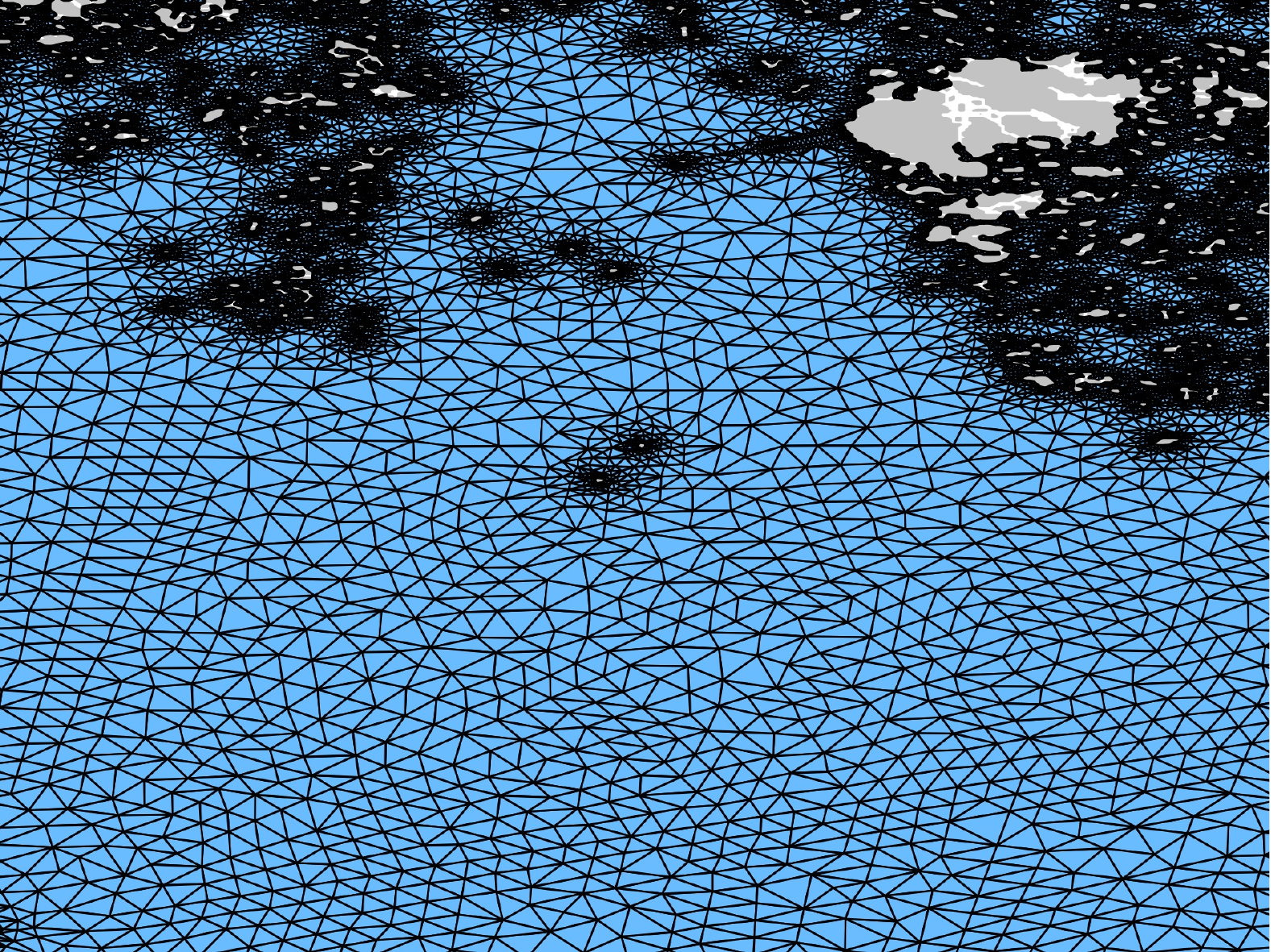}}&
    \setlength{\fboxsep}{0pt} \fbox{\includegraphics[width=7cm]{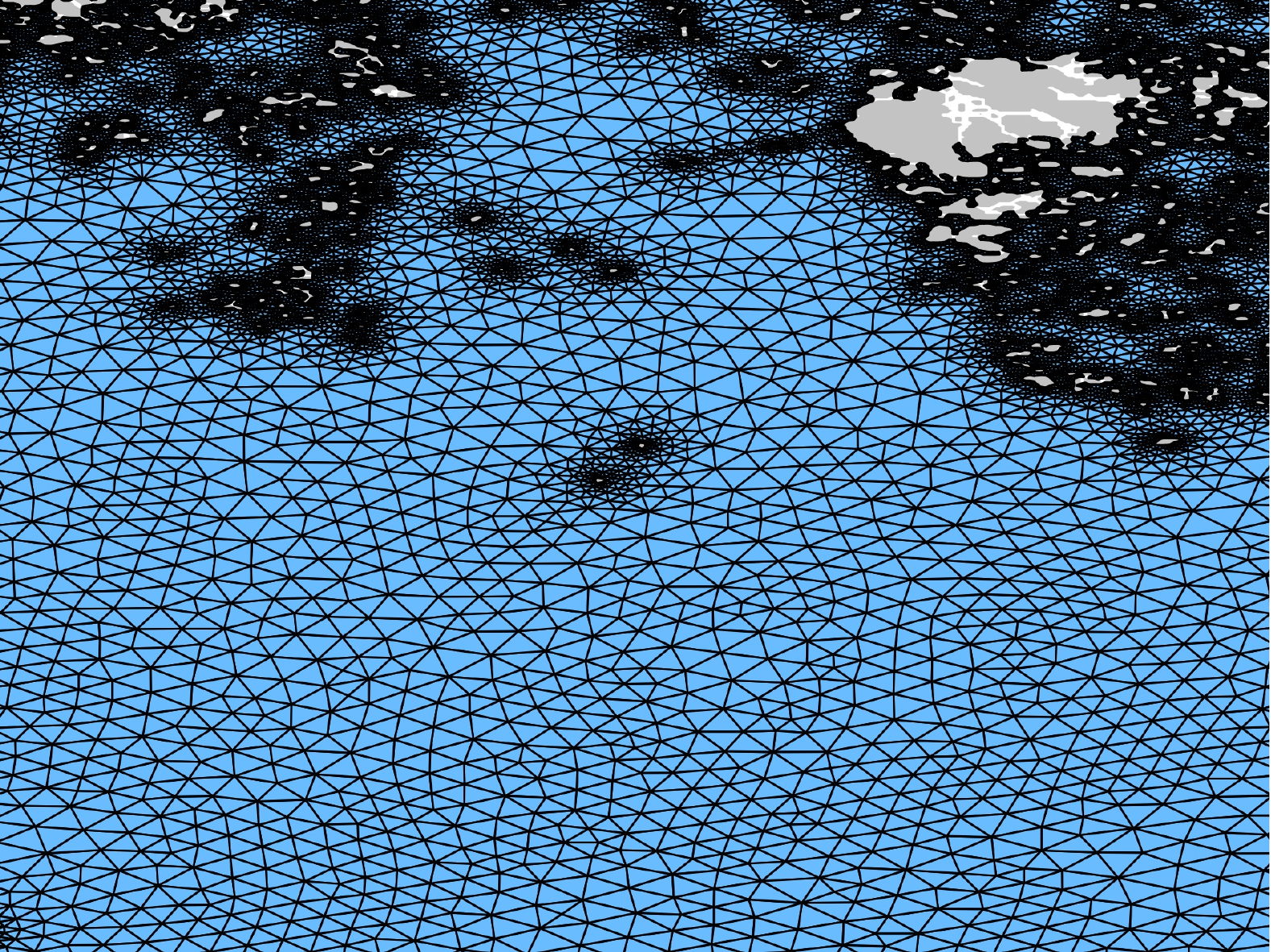}}\\
    \footnotesize{8 refinement steps (988,066 points)}&
    \footnotesize{8 refinement steps and smoothing}
  \end{tabular}
\caption{Saturation of the edges.
  The final mesh (iteration 8) contains 2,345,823 triangles.
\label{fig:balt2}}
  \end{center}
\end{figure}
 
\begin{figure}[h!]
  \begin{center}
 \begin{tabular}{c}
   \setlength{\fboxsep}{0pt} \fbox{\includegraphics[height=7cm]{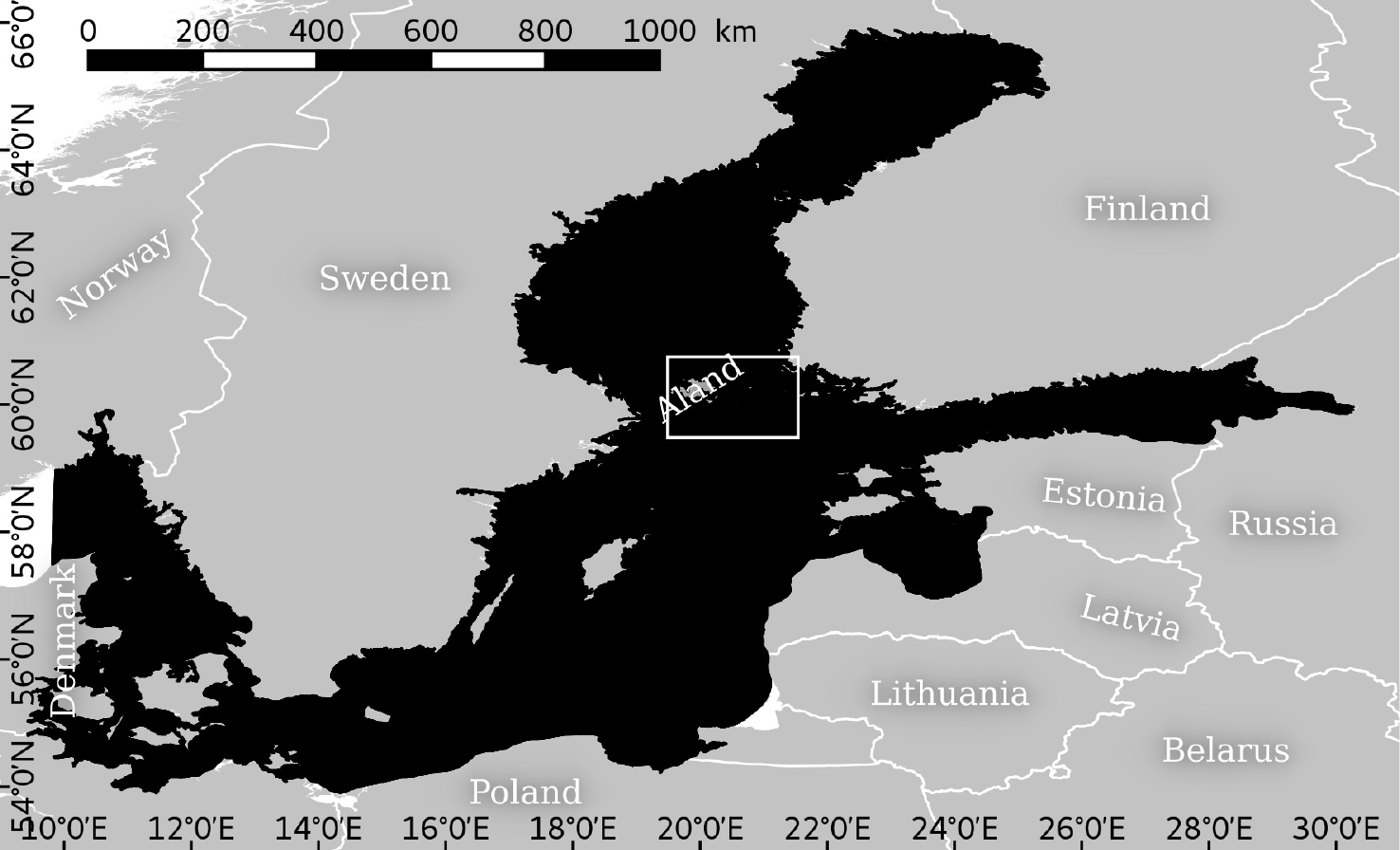}}\\
   \setlength{\fboxsep}{0pt} \fbox{\includegraphics[height=7cm]{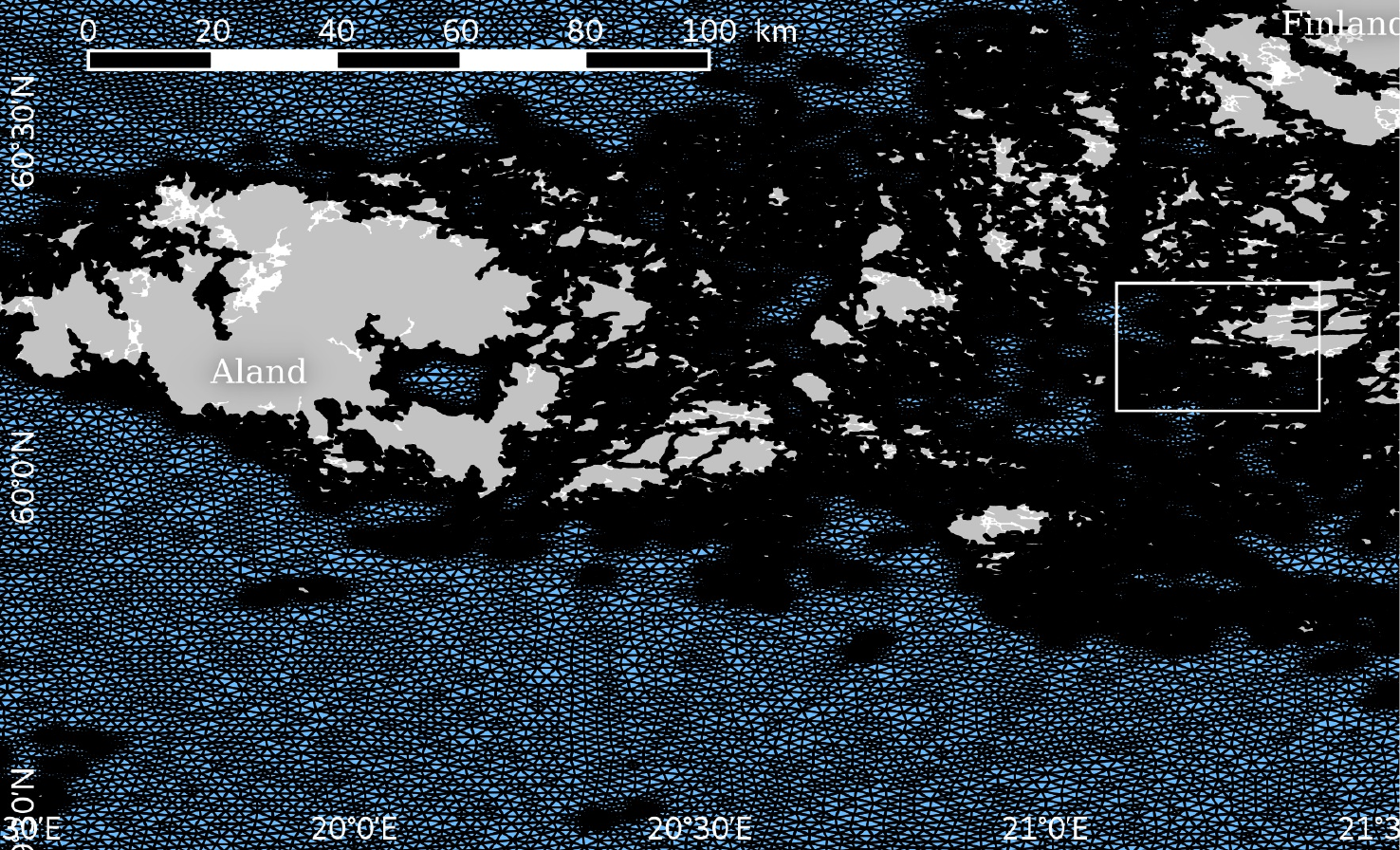}}\\
   \setlength{\fboxsep}{0pt} \fbox{\includegraphics[height=7cm]{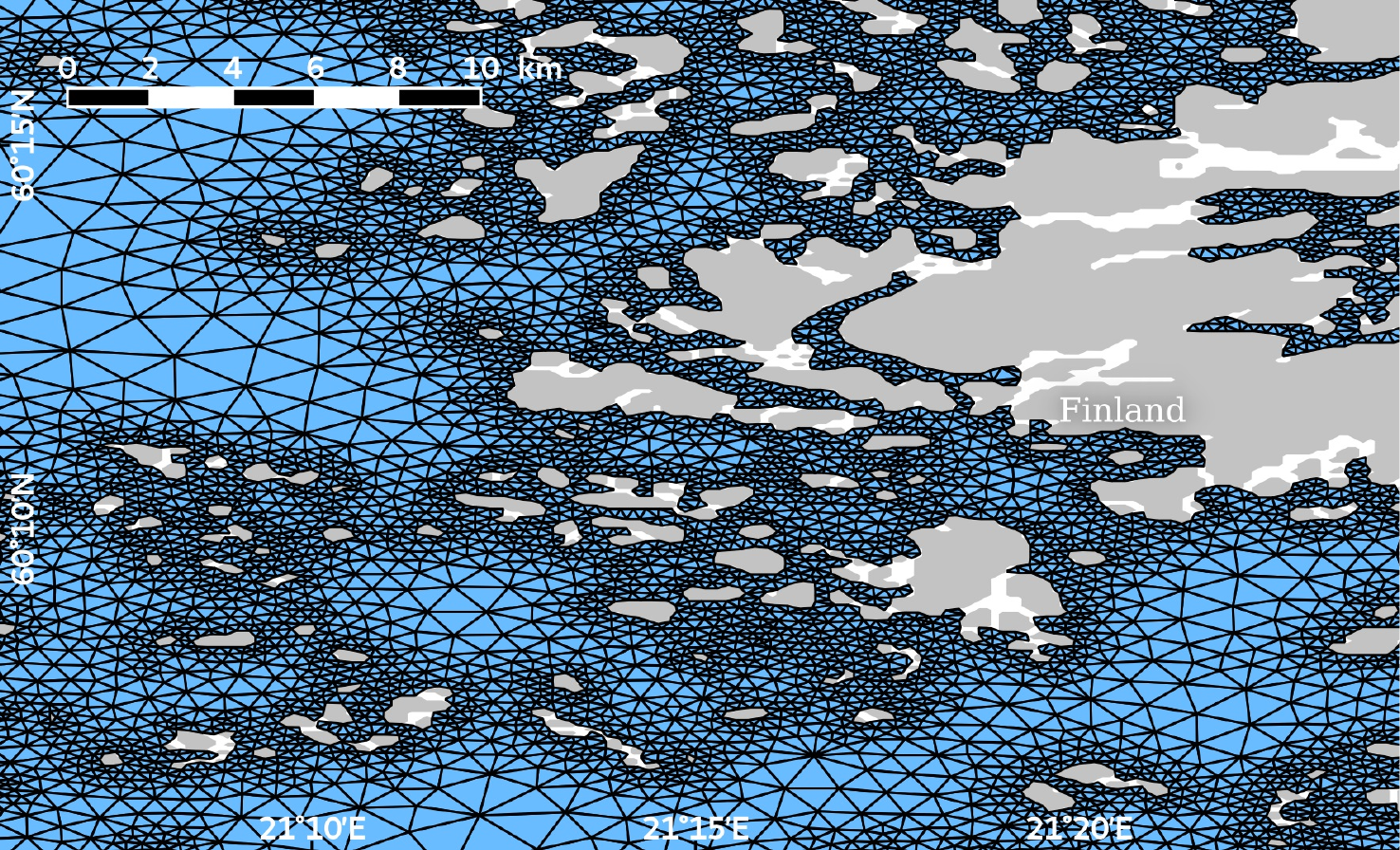}}
\end{tabular}
\caption{Images of the final mesh of the Baltic sea.
\label{fig:balt3}}
  \end{center}
\end{figure}

 At stage $i$ of the refinement process, all internal edges of the
domain  are saturated, producing a point set
$\pointSet_i$. Points are spaced rightfully on edges of the existing
mesh. Yet, there is no guarantee that points belonging to different
edges are not too close to each other. Here, some filtering is
required. Two approaches for filtering the points have been
investigated. At first,  
a space search structure has been used (a RTree \cite{beckmann1990r}). Using such a
structure involves extra-storage. Insertions on a RTree in parallel is
not an easy matter as well.

A more straightforward approach has been developed. The Delaunay kernel is 
applied to all points $p_k$ of $\pointSet_j$ without filtering. 
Consider triangulation ${\DT}_{k-1}$ and the Delaunay cavity 
$ {\mathcal C}({\DT}_{k-1},{p}_{k})$ relative to point $p_k$ that has
to be inserted in the mesh.
There exist no points of ${\DT}_{k-1}$ that are closer to $p_k$ than the
points of ${\mathcal C}({\DT}_{k-1},{p}_{k})$ thanks to the Delaunay
property.  Point $p_k$ is inserted if and only if no short edge is 
created during the insertion. It is thus only necessary to check
points of the cavity in the filtering process.

In our implementation, both approaches have shown to provide similar serial
performances. Yet, the approach based on the Delaunay kernel is
clearly advantageous: (i) it is way easier to code ({\it ceteris
  paribusis}, a solution that is simpler to code is always better),
(ii) it does not require any overhead and, (iii) more important, 
it enjoys the multi-threaded implementation of the Delaunay kernel.

Figure \ref{fig:balt2} illustrates the Delaunay refinement procedure
on the Baltic sea. Final mesh of $2,345,823$ triangles has been
generated in $16$ seconds on one thread. The size field function $h$ that has been chosen is
highly computational intensive: The Delaunay kernel takes only $5,44$
seconds. Sorting the points along a 3D Hilbert curve takes $1,29$
seconds and saturating the edges takes $9,25$ seconds which is way more
than half of the total CPU time. Note that parallelizing this stage of
the algorithm is actually trivial. The algorithm converges when every
edge has an adimensional length less or equal to oen. Eight iterations were necessary for
convergence. Most of the points were inserted during the two first
iterations (see Figure \ref{fig:balt2}). Images of the final mesh are
shown if Figure \ref{fig:balt3}.

\section{World ocean}
\begin{figure}[t!]
  \begin{center}
 \begin{tabular}{cc}
   \setlength{\fboxsep}{0pt} \fbox{\includegraphics[height=7cm]{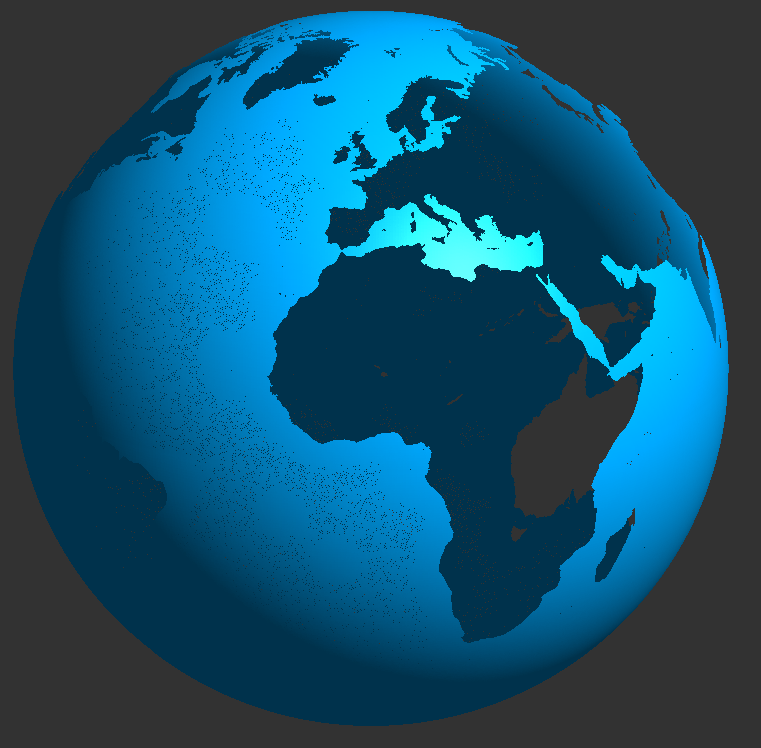}}&
   \setlength{\fboxsep}{0pt} \fbox{\includegraphics[height=7cm]{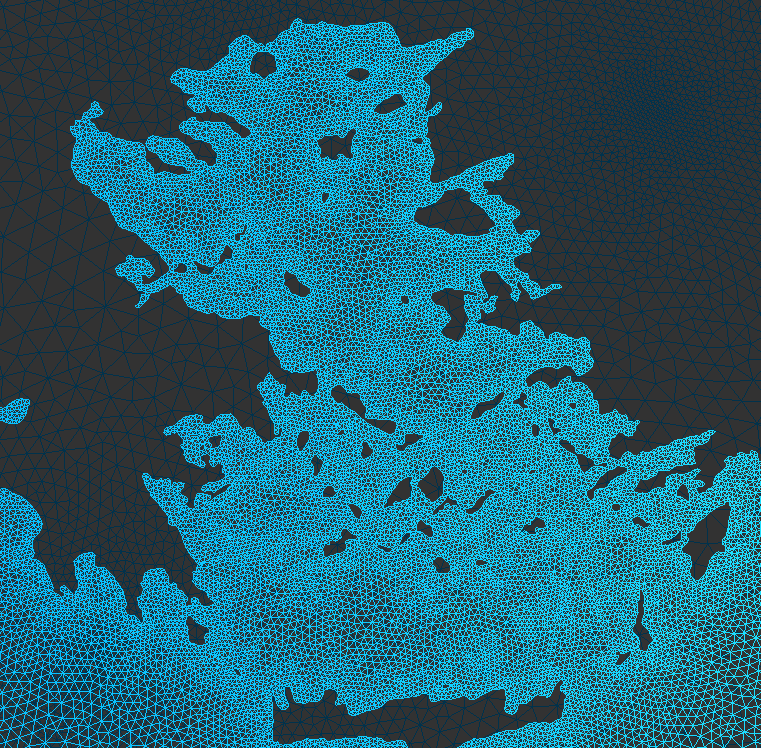}}\\
\end{tabular}
   \setlength{\fboxsep}{0pt} \fbox{\includegraphics[height=8.1cm]{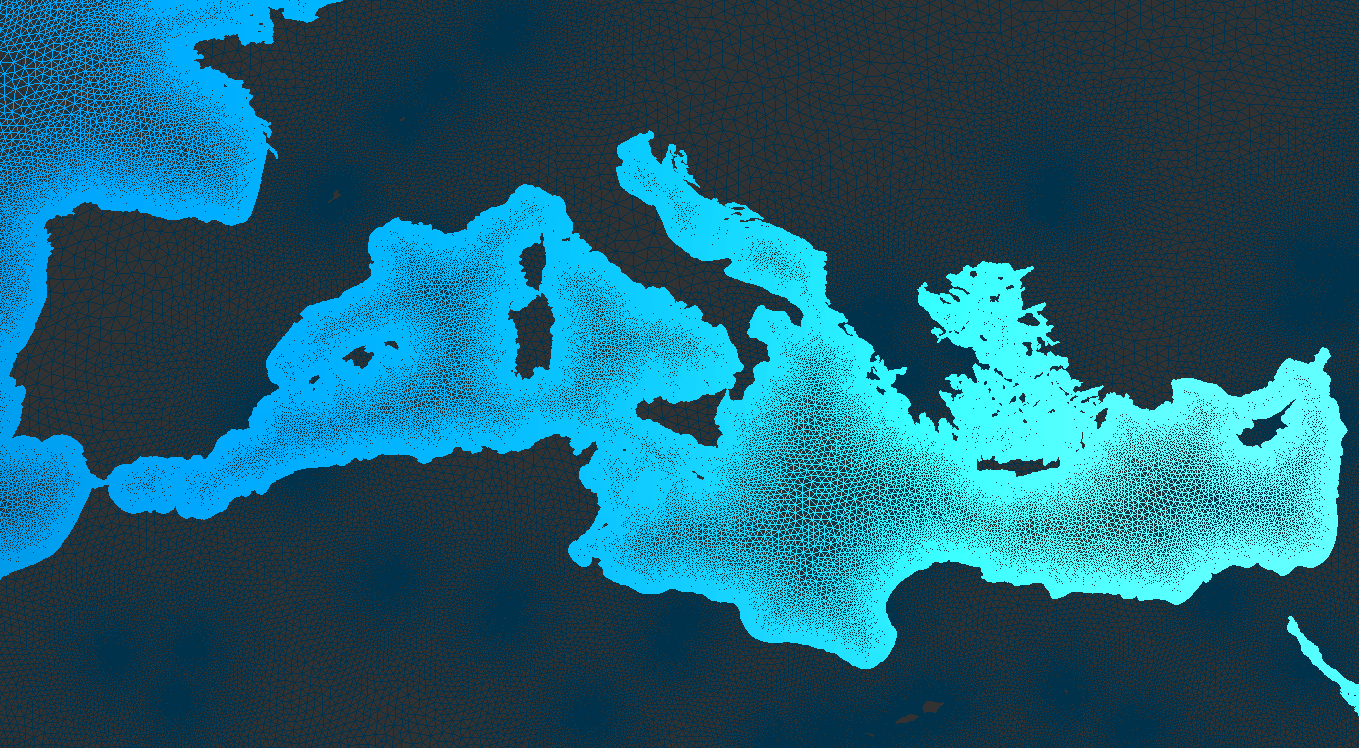}}
\caption{Images of the mesh of the world ocean (top left) with a zoom on the
  Mediterranean sea (bottom)  and on the Egean sea (top right).
\label{fig:world}}
  \end{center}
\end{figure}

In this section, we have generated a mesh of the whole world ocean
with a global resolution of $h_{\min} = 3$km. A larger mesh size of 
$h_{\min} = 30$km was chosen far away from the coast like in 
Equation \eqref{eq:size} (Figure \ref{fig:world}). The final mesh
contains 4,978,243 triangles and has been generated in about 50
seconds. About 30\% of the total time was taken for generating the
initial Delaunay mesh of the data. In the Delaunay refinement procedure
14.5 seconds were necessary to generate the points (saturation of the
edges), 2.76 seconds were necessary to sort point sets $\pointSet_j$
and 8.44 seconds were used in the Delaunay kernel.

\section{Conclusions}

This paper presents an original algorithm that allow to generate
meshes of domains defined by coastline datas on the sphere. The main
novelties of the approach are (i) the treatment of input data, (ii)
the multithreaded nature of the algorithms and (iii) the Delaunay
kernel on the sphere with automatic filtering. The extension to
quadrilateral meshing is the following step in our developments. 

The code that has been used here is a self consistent piece of code
that will be released soon as an open source. It will be part of the
Gmsh distribution.

Parallel computations will be presented in further communications.

\bibliographystyle{abbrv}
\bibliography{biblio}

\end{document}